\documentclass[11pt,a4paper,onecolumn]{IEEEtran}

\usepackage{graphicx}
\usepackage{amsmath, amsthm, amssymb, amsfonts, cancel}
\usepackage{algorithm, algorithmic, ifsym}

\newcommand\blfootnote[1]{%
  \begingroup
  \renewcommand\thefootnote{}\footnote{#1}%
  \addtocounter{footnote}{-1}%
  \endgroup
}

\def\Cov{{\rm Cov}}
\def\tr{{\rm tr}}

\def\bp{ \textbf{p} } 
\def\ba{ \textbf{a} } 
\def\bA{ {\mathbf{A}} }

\def\bY{ {\mathbf{Y}} }

\def\bepsilon{ {\mathbf{\epsilon}} }

\def\nn{{ \parallel   }}
\def\RR{{ \mathbb{R}  }}
\def\PP{{ \mathbb{P}  }}
\def\EE{{ \mathbb{E}  }}
\def\NN{{ \mathbb{N}  }}

\def\tr{{ \text{tr}   }}

\def\by{{ \mathbf{y}  }}

\def\bg{ \mathbf{g}  }
\def\card{{ \text{card} }}

\newtheorem{theorem}{Theorem}
\newtheorem{lemma}{Lemma}

\newtheorem{corollary}{Corollary}

\newtheorem{assumption}{Assumption}

\begin{document}
\title{Collaborative 20 Questions for Target Localization}
\author{Theodoros Tsiligkaridis *, \textit{Student Member, IEEE}, Brian M. Sadler, \textit{Fellow, IEEE},\\ Alfred O. Hero III, \textit{Fellow, IEEE}}

\maketitle

\begin{abstract} 
We consider the problem of 20 questions with noise for multiple players under the minimum entropy criterion \cite{Jedynak12} in the setting of stochastic search, with application to target localization. Each player yields a noisy response to a binary query governed by a certain error probability. First, we propose a sequential policy for constructing questions that queries each player in sequence and refines the posterior of the target location. Second, we consider a joint policy that asks all players questions in parallel at each time instant and characterize the structure of the optimal policy for constructing the sequence of questions. This generalizes the single player probabilistic bisection method \cite{Jedynak12, CastroNowak07} for stochastic search problems. Third, we prove an equivalence between the two schemes showing that, despite the fact that the sequential scheme has access to a more refined filtration, the joint scheme performs just as well on average. Fourth, we establish convergence rates of the mean-square error (MSE) and derive error exponents. Lastly, we obtain an extension to the case of unknown error probabilities. This framework provides a mathematical model for incorporating a human in the loop for active machine learning systems.
\end{abstract}


\blfootnote{The research reported in this paper was supported in part by ARO grant W911NF-11-1-0391. Preliminary results in this paper have appeared at the 2013 IEEE International Conference on Acoustics, Speech and Signal Processing and at the 2013 IEEE GlobalSIP - Symposium on Controlled Sensing For Inference: Applications, Theory and Algorithms.

T. Tsiligkaridis was with the Department of Electrical Engineering and Computer Science, University of Michigan, Ann Arbor, MI 48109 USA. He is now with MIT Lincoln Laboratory, Lexington, MA 02421 USA (email: ttsili@ll.mit.edu).

B. M. Sadler is with the US Army Research Laboratory, Adelphi, MD 20783 USA (email: brian.m.sadler6.civ@mail.mil).

A. O. Hero is with the Department of Electrical Engineering and Computer Science, University of Michigan, Ann Arbor, MI 48109 USA (email: hero@umich.edu).

}

\section{Introduction} \label{sec:intro}
What is the intrinsic value of adding another sensor in a network performing sequential estimation of a target driven by active queries? How can two or more experts communicating over noisy binary symmetric channels best collaborate to localize a target when the channel crossover probabilities are unknown? A simple model for answering these questions is a collaborative multi-player 20 questions game, where the players (sensors or experts) in the network are repeatedly queried about the location of an unknown target $X^*$ in order to improve estimation performance. This paper proposes such a 20 questions framework for sequentially optimizing the queries in a general setting that can handle multiple players (sensors) with different levels of accuracy, which can be time varying, and different costs of querying.


Motivated by the approach of Jedynak, et al., \cite{Jedynak12}, which was restricted to the single player case, we model the player interactions as a noisy  collaborative 20 questions game. In this framework a controller sequentially selects a set of questions about target location and uses the noisy responses of the players to formulate the next set of questions. 
Under flexible noisy query-response models for errors in the player's responses, we derive the optimal query policy, establish an equivalence theorem, and obtain tight performance bounds. We illustrate the flexibility of our general framework by specializing the theory to a simple human-machine interaction model, incorporating a time varying error model previously proposed for modeling human response \cite{Jamieson12}. 
The query response models assumed for the human and the machine are different, but complementary. While the machine's accuracy is constant over time, the target localization accuracy of the human degrades over time (as in the derivative-free optimizers (DFO) model of Jamieson et al \cite{Jamieson12}), reflecting the human's decreased ability to resolve questions about the precise target location near the end of the game. Our model predicts that the value of including the human-in-the-loop (as measured by a quantity called the human gain ratio (HGR)) initially increases when localization errors are large, and then slowly decreases over time as the location errors go below the human's fine resolution capability.

While not pursued in this paper, another possible application of our collaborative 20 questions framework is crowdsourcing. Crowdsourcing systems distribute a large number of tasks to many workers in order to efficiently solve large-scale data-processing tasks in various domains. Under a fairly general model of crowdsourcing tasks, the problem of minizing the number of task assignments subject to a constraint on the overall reliability is studied in \cite{Karger:2013}. Karger et al. \cite{Karger:2013} propose a non-adaptive algorithm for deciding which tasks to allocate to which workers. The algorithm also tries to infer the correct answers given the noisy workers' responses. In \cite{Karger:2013}, it is shown that adaptive and non-adaptive allocation approaches behave similarly in terms of their ability to optimize the task assignment to workers subject to the worker reliability constraints. Our framework might be used in the design of crowdsourcing systems for target search problems to optimize the questions assigned to workers that operate with different levels of accuracy (and may be unknown).  



The roots of optimal query design lie in stochastic control \cite{Puterman, Bertsekas:1996}. Applications of this methodology include active learning \cite{Settles:2009, CastroNowak06, CastroPhD, CastroNowak07} and sequential experimental design \cite{DeGroot:1970, Wetherill:1986}. For Bayesian formulations it is known that the Bayes-optimal policy that arises is the solution to a partially observed Markov decision process (POMDP), which is described by a dynamic programming recursion. While it is sometimes possible to obtain explicit solutions to this recursion \cite{GittinsJones:1974, Berry:1985}, in many cases it is intractable. As a result, when the globally optimal policy is too difficult to compute, a one-step lookahead heuristic is often used as a greedy approximation \cite{Zhang:2003}.

A key motivator for our work is the paper by Jedynak et al \cite{Jedynak12}, where a Bayesian formulation is considered for sequential estimation of the target location. The problem was formulated in the context of a 20 questions game and it was shown that the greedy policy is Bayes-optimal under a minimum expected entropy criterion. In addition, under a noisy response and a symmetric noise model, bisecting the posterior yields globally optimal policies after a finite number of questions. This posterior bisection policy has been called the probabilistic bisection algorithm (PBA), or Horstein's scheme, and has roots in information theory \cite{Horstein:1963} in the context of sequential encoding of a message through a binary symmetric channel (BSC). The origins of the entropy minimizing 20 questions game lie in information theoretic formulations of binary search \cite{CoverThomas}. The binary search procedure was further studied in \cite{NowakGBS:2011}, where under incoherence conditions, the generalized binary search (GBS) can learn a ``correct" binary-valued function through a sequence of $O(\log N)$ queries in a space of $N$ hypothesized functions. This method has also been applied to the problem of learning halfspaces in machine learning. Another related work is Hegedus's halving algorithm \cite{Hegedus:1995} that attempts to identify an unknown target concept $c^*$ chosen from a known concept class $\mathcal{C}$, making queries about $c^*$. It was shown in \cite{Hegedus:1995} that any target concept can be identified in at most $\left\lfloor {\log |\mathcal{C}|} \right\rfloor$ queries. A variant of the 20 questions problem for target search was also studied in the context of noisy comparison trees \cite{Feige:1994}, where query complexity bounds were derived for various search-related problems, including binary search, sorting and merging.

Another problem related to target search is stochastic root-finding. In this problem, the target is the zero of a decreasing function $f$, and the task is to locate the root of $f$ given noisy observations of the function. The controller chooses the query points $x_1,x_2,...$ and observes noisy versions of $f(x_1),f(x_2),...$. The queries in this setting are questions of the form ``Is $f(x) < 0$?", and rates of convergence are well known. In \cite{Waeber:2011}, it was shown that under mild conditions on the noisy response models, a probabilistic bisection method converges to the root of $f$ almost surely. In addition, for the constant error rate case, it was also shown that it converges exponentially fast; contrary to the best stochastic approximation rate of $n^{-1/2}$ \cite{RobbinsMonro:1951, Kushner:2003}.

\subsection{Contributions}
All the aformentioned works consider the single player case-i.e., a single query is designed at each time instant and a single noisy response on the target's
location is obtained. 
In this paper, we consider the collaborative multiplayer case and derive corresponding optimality conditions for optimal query strategies when there is no restriction on query complexity. We propose a sequential bisection policy for which each player responds to a single question about the location of the target, and a joint policy where all players are asked questions simultaneously. We show that even when the collaborative players act independently, jointly optimal policies require overlapping non-identical queries. We prove that the maximum entropy reduction for the sequential bisection scheme is the same as that of the jointly optimal scheme, and is given by the sum of the capacities of all the players' channels. This is important since, while the jointly optimal scheme might be hard to implement as the number of players and dimensions increase, the sequential scheme only requires a sequence of bisections followed by intermediate posterior updates. Thus, by implementing the sequential policy, complexity is transferred from the controller to the posterior updates. Despite the fact that the optimal sequential policy has access to a more refined filtration, it achieves the same average performance as the optimal joint policy. An anonymous reviewer pointed out that this equivalence result is good news for applications where it is impractical to perform sequential queries, e.g., in experimental biology where multiple experiments are most easily performed in a batch instead of sequentially.

We also extend the results to the case where there are costs to querying different players. Specifically, we consider the player selection problem in addition to the query design problem and strike a balance between uncertainty reduction on the target's location and cost of each player. This is important in practice.  For example, it may be that the more accurate a player is, the higher the cost of use.

We extend this equivalence of jointly designed and sequentially designed queries to the setting where the error channels associated with the players are unknown. In this case, we show that the expected entropy loss at each iteration is no longer constant; it is time-varying and equals the conditional expectation of the sum of the capacities of the players' channels with respect to the filtration up to the current time. In addition, we show that even for one-dimensional targets, the optimal policy for the unknown channel case is not equivalent to the probabilistic bisection policy. 

The work by Castro and Nowak \cite{CastroNowak07, CastroNowak06} provides upper bounds on the MSE of the median of the posterior distribution of the target for the single player case. We extend their MSE bounds to the multiplayer case and provide new lower bounds on MSE by linking our information theoretic analysis to convergence rates. The combination of the upper and lower bounds sandwiches the MSE between two exponentially decaying functions of the number of plays in the 20 questions game.

Our 20 questions framework bears some similarity to other binary forced choice problems that have appeared in the literature. This includes educational testing, e.g., using dynamic item response models \cite{Wang:2012}, and active learning, e.g., using paired comparisons for ranking two objects \cite{Jamieson:2011}. Like the 20 questions framework, in \cite{Wang:2012, Jamieson:2011}, a sequence of binary questions is formulated by a controller. However, the 20 questions problem considered in this paper is quite different. The goals are not the same: in contrast to sequential testing considered in \cite{Wang:2012, Jamieson:2011}, here as in \cite{Jedynak12} we consider sequential estimation of a continuous valued target state. Furthermore, in \cite{Wang:2012, Jamieson:2011} the queries are posed to a single player whereas we consider multiple players who cooperate to accomplish posterior entropy minimization.

\subsection{Outline}
The outline of this paper is as follows. Section \ref{sec:single_player} provides background and introduces some notation for the 20 questions problem. Section \ref{sec: multiplayer_entropy_loss} introduces the collaborative player setup. It introduces the sequential bisection policy and the joint policy, and establishes that the respective optimal policies attain identical performance. Section \ref{sec: bound_MSE} derives performance bounds on the MSE and Section \ref{sec: human_in_the_loop} develops similar bounds for a human error model. 
Section \ref{sec: unknown_error_probability} extends the analysis to the case when the error probabilities are not known. The theory is illustrated by simulation in Section \ref{sec: simulations} and is followed by our conclusions in Section \ref{sec: conclusions}. 

\section{Noisy 20 Questions with a Single Player} \label{sec:single_player}
Jedynak, et al. \cite{Jedynak12} formulate the single player 20 questions problem as follows. A controller queries a noisy oracle about whether or not a target $X^*$ lies in a measurable set $A_n \subset \RR^d$.\footnote{For technical reasons, this is taken to be the union of at most $J_n$ half-open intervals in the problem formulation of \cite{Jedynak12}.} At time $n$, the noisy response $Y_{n+1}$ is a probabilistic function of the indicator function $Z_n=I(X^*\in A_n)$ and the mapping is modeled as a binary memoryless and time-invariant channel. Each query-response pair $(A_n,Y_{n+1})$ transforms the posterior distribution from $p_n(\cdot)$ to $p_{n+1}(\cdot)$. Starting with a prior distribution on the target's location $p_0(\cdot)$, the objective in \cite{Jedynak12} is to minimize the expected entropy of the posterior distribution $p_N(\cdot)$ after having asked $N$ questions:
\begin{equation} \label{eq: min_entr_obj}
	\inf_\zeta \EE^\zeta \left[ H(p_N) \right],
\end{equation}
where $\zeta=(\zeta_0,\zeta_1,\dots)$ denotes the controller's query policy and $\EE^\zeta[\cdot]$ denotes the expectation taken with respect to the probability measure on $\{A_n,Y_{n+1}\}_{n=0}^{N-1}$ induced by the policy $\zeta$. The entropy is the standard differential entropy \cite{CoverThomas}:
\begin{equation*}
	H(p) = -\int_{\mathcal{X}} p(x) \log p(x) dx.
\end{equation*}
The median of the posterior distribution $p_N(\cdot)$ is used to estimate the target location after $N$ questions. Jedynak \cite{Jedynak12} shows the bisection policy is optimal under the minimum entropy criterion. To be concrete, in Theorem 2 of \cite{Jedynak12}, optimal policies are characterized by:
\begin{equation} \label{eq: optimal_policy}
	\PP_n(A_n) := \int_{A_n} p_n(x) dx = u^* \in \arg \max_{u \in [0,1]} \phi(u),
\end{equation}
where
\begin{equation*}
	\phi(u) = H(f_1 u + (1-u) f_0) - u H(f_1) - (1-u) H(f_0)
\end{equation*}
is nonnegative. The densities $f_0$ and $f_1$ correspond to the noisy channel \footnote{The function $I(A)$ is the indicator function throughout the paper-i.e., $I(A)=1$ if $A$ is true and zero otherwise.}:
\begin{equation*}
	\PP(Y_{n+1}=y| Z_n=z) = f_0(y) I(z=0) + f_1(y) I(z=1),
\end{equation*}
where $Z_n=I(X^*\in A_n) \in \{0,1\}$ is the channel input. While the framework applies to both continuous and discrete random variables $y$, in \cite{Jedynak12} the focus was on the binary case-i.e., $y\in\{0,1\}$. The noisy channel models the conditional probability of the response to each question being correct. For the special case of a binary symmetric channel (BSC), in (\ref{eq: optimal_policy}) $u^* = 1/2$ and the probabilistic bisection policy \cite{Jedynak12, CastroNowak07} becomes an optimal policy.

\section{Noisy 20 Questions with Collaborative Players: Known Error Probability} \label{sec: multiplayer_entropy_loss}

Assume that there is a target with unknown state $X^*\in \mathcal{X} \subset \RR^d$. We focus on the case where the target state is spatial location, i.e., in $d=2$ or $3$ dimensions. However, our results are applicable to higher dimensions also, e.g., where $X^*$ is a kinematic state or some other multi-dimensional target feature. Starting with a prior distribution $p_0(\cdot)$ on $X^*$, the aim is to find an optimal policy for querying a set of players about the target state. The policy's objective is to minimize the expected Shannon entropy of the posterior distribution $p_n(\cdot)$ of the target location after $n$ questions.

There are $M$ collaborating players that can be asked questions at each time instant $n$. The objective of the players is to come up with the correct answer to a kind of 20 questions game. Next, we introduce two types of query design strategies. The first is a sequential strategy where the controller formulates and asks questions to each player in sequence. The second is a batch strategy where the questions are formulated and directed to all players simultaneously. For fixed $n$ both strategies ask the same number of questions. However, the sequential strategy has the advantage of being able to use the answer of the previous player to better formulate a question to the next one. Below we show that, despite this advantage, the average entropy reduction performances of these two strategies are identical.

\subsection{Sequential Query Design}

\begin{figure}[t]
	\centering
		\includegraphics[width=0.40\textwidth]{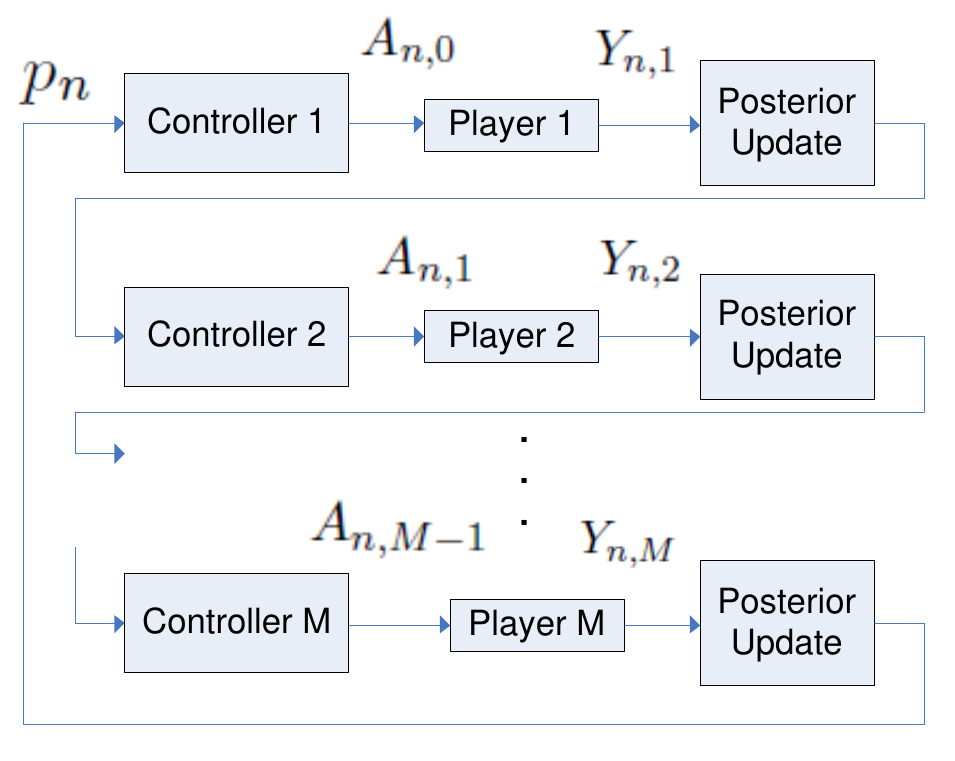}
	\caption{\small{Controllers sequentially ask questions to $M$ collaborative players about the location $X^*$ of an unknown target. At time $n$, the first controller chooses the query $I(X^*\in A_{n,0})$ based on the posterior $p_n$. Then, player 1 yields the noisy response $Y_{n,1}$ that is used to update the posterior, and the second controller chooses the next query $I(X^* \in A_{n,1})$ for player 2 based on the updated posterior, and so on.} }
	\label{fig:seq_scheme}
\end{figure}

The sequential strategy is the following coordinate-by-coordinate design: ask an optimal query to the first player, then update the posterior density and ask an optimal query to the second player, and so on (see Figure \ref{fig:seq_scheme}). In \cite{Jedynak12}, the optimal query policy for the case of a single player ($M$=1) was shown to be a bisection rule. 

For each time epoch, indexed by $n$ and called a cycle, the controller formulates and asks  the $M$ players questions  $A_{n_t}=A_{n,t}$, $t=0, \ldots, M-1$. We denote by $n_t=(n,t)$  the times at which the queries are asked.

Let the $m$th player's query at time $n_{t}=n_{m-1}$ be ``does $X^*$ lie in the region $A_{n_t} \subset \RR^d$?''. We denote the truth state of the query as the binary variable $Z_{n_t}=I(X^*\in A_{n_t}) \in \{0,1\}$ and the noisy binary response of the $m$th player is $Y_{n_{t+1}} \in \{0,1\}$. 

The query region $A_{n_t}$ chosen at time $n_t$ depends on the information available at that time. More formally, define the multi-index $(n,t)$ where $n=0,1, \ldots$ indexes over cycles and $t=0, \ldots, M-1$ indexes within cycles.  Define the nested sequence of sigma-algebras $\mathcal G_{n,t}$, $\mathcal  G_{n,t} \subset \mathcal G_{n+i,t+j}$, for all $i\geq 0$ and $j\in\{0,\dots,M-1-t\}$, generated by the sequence of queries and the players' responses. The filtration $\mathcal G_{n,t}$ carries all the information accumulated by the controller from time $(0,0)$ to time $(n,t)$.  The queries $\{A_{n,t}\}$ formulated by the controller are measurable with respect to this filtration.


\subsection{Joint Query Design} \label{subsec: joint_scheme}

\begin{figure}[t]
	\centering
		\includegraphics[width=0.50\textwidth]{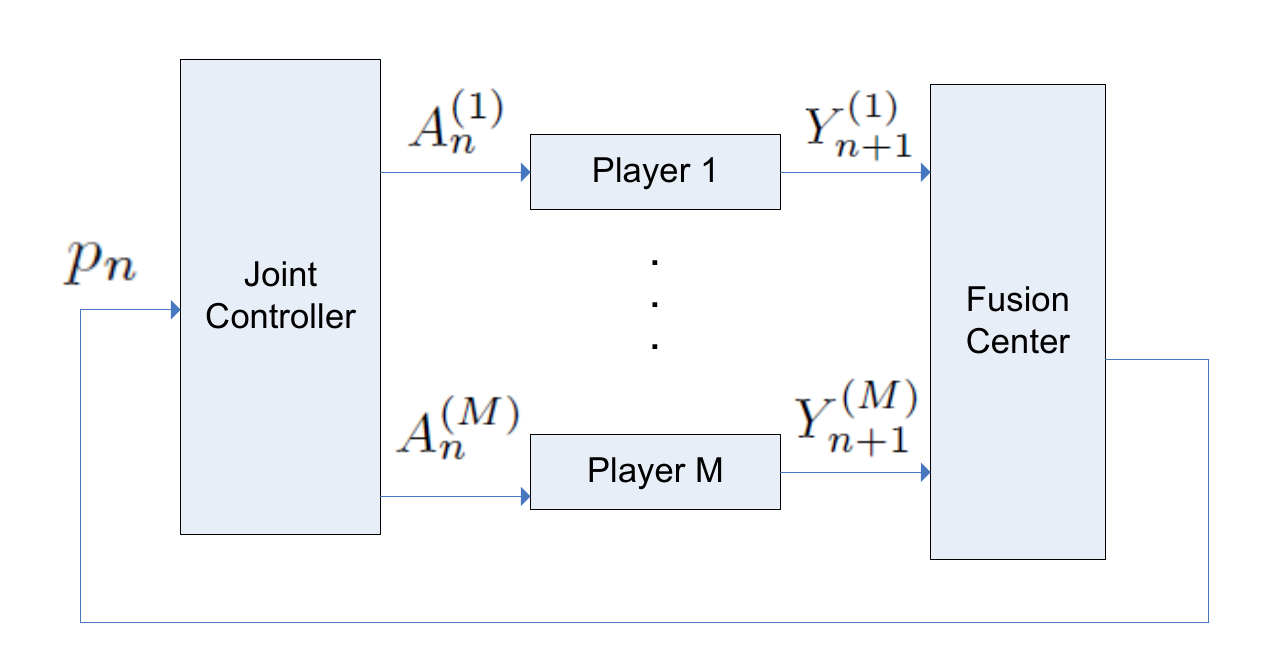}
	\caption{\small{A controller asks a batch questions of $M$ collaborative players about the location $X^*$ of an unknown target. At time $n$, the controller chooses the queries $I(X^*\in A_n^{(m)})$ based on the posterior $p_n$. Then, the $M$ players yield noisy responses $Y_{n+1}^{(m)}$ that are fed into the fusion center, where the posterior is updated and fed back to the controller at the next time instant $n+1$. } }
	\label{fig:joint_scheme}
\end{figure}

Let the $m$th player's query at time $n$ be ``does $X^*$ lie in the region $A_n^{(m)} \subset \RR^d$?''. We denote this query as the binary variable $Z_n^{(m)}=I(X^*\in A_n ^{(m)}) \in \{0,1\}$ to which the player provides a possibly incorrect (i.e., noisy) binary response $Y_{n+1}^{(m)} \in \{0,1\}$. We consider a similar setting as in \cite{Jedynak12}, which applied to the $M=1$ player case, but now we have a joint controller that chooses a batch of $M$ queries $\{A_n^{(m)}\}_{m=1}^M$ that are addressed to each of the $M$ players at time $n$. A block diagram is shown in Figure \ref{fig:joint_scheme}.

As in the sequential query design, the joint queries are selected based on the accumulated information available to the controller. However, since the full batch of joint queries are determined at the beginning of the $n$-th cycle, the joint controller only has access to a coarser filtration $\mathcal{F}_n$, $\mathcal{F}_{n-1} \subset \mathcal{F}_{n} $, as compared with the filtration $\mathcal{G}_{n,t}$ of the sequential controller.


\subsection{Definitions \& Assumptions}
We next present a set of assumptions and definitions that will be used throughout the paper. 

Define the $M$-tuples $\bY_{n+1}=(Y_{n+1}^{(1)},\dots,Y_{n+1}^{(M)})$ and $\bA_n = \{A_n^{(1)},\dots,A_n^{(M)}\}$. 

\begin{assumption} (Conditional Independence) \label{assump:cond_indep}
We assume that the players' responses are conditionally independent. In particular, for the joint controller,
\begin{align}
	\PP(&\bY_{n+1}=\by|\bA_n,X^*=x,\mathcal{F}_n) \nonumber \\
		&= \prod_{m=1}^M \PP(Y_{n+1}^{(m)}=y^{(m)}|A_n^{(m)},X^*=x,\mathcal{F}_n), \label{eq:indep}
\end{align}
where
\begin{align}
	\PP(&Y_{n+1}^{(m)}=y^{(m)}|A_n^{(m)},X^*=x,\mathcal{F}_n) \nonumber \\
		&= \Bigg\{ \begin{array}{ll} f_1^{(m)}(y^{(m)}|A_n^{(m)},\mathcal{F}_n), & x\in A_n^{(m)} \\ f_0^{(m)}(y^{(m)}|A_n^{(m)},\mathcal{F}_n), & x\notin A_n^{(m)} \end{array}. \label{eq:exp}
\end{align}
Similar relations hold for the sequential controller under the conditional independence assumption: in (\ref{eq:indep}) and (\ref{eq:exp}) simply change the subscripts $n$ and $n+1$ to $n_t$ and $n_{t+1}$, respectively, and replace the filtration $\mathcal{F}_n$ by $\mathcal{G}_{n,t}$. 
\end{assumption}

\begin{assumption} (Memoryless Binary Symmetric Channels) \label{assump:BSC}
	We model the players' responses as independent (memoryless) binary symmetric channels (BSC) \cite{CoverThomas} with crossover probabilities $\epsilon_m\in (0,1/2)$. In particular, for the joint query strategy, the conditional probability mass function $f_j^{(m)}=\PP(Y^{(m)}_n=j|A_n^{(m)},\mathcal{F}_n)$ of the response of the $m$-th player is:
\begin{align*}
	f_j^{(m)}&(y^{(m)}|A_n^{(m)},\mathcal{F}_n)=f_j^{(m)}(y^{(m)}) \\
		&= \Bigg\{ \begin{array}{ll} 1-\epsilon_m, & y^{(m)}=j \\ \epsilon_m, & y^{(m)}\neq j \end{array}
\end{align*}
where $m=1,\dots,M, j=0,1$. We note that the channel may depend on the posterior distribution $p_n(\cdot)$ (or any function of it). A similar relation holds for the sequential query strategy: replace $n$ by $n_t$ and $\mathcal{F}_n$ by $\mathcal{G}_{n,t}$.
\end{assumption}

\begin{assumption} (Query Region Regularity) \label{assump:query_region_regularity}
	We restrict the query region $A_n^{(m)}$ for each player $m$ to be the union of at most $J_n^{(m)}$ rectangles, i.e., 
	\begin{equation*}
		A_n^{(m)} = \bigcup_{j=1}^{J_n} R_n^{(m)}(j)
	\end{equation*}
	where $R_n^{(m)}(j)=[a_n^{(m)}(j;1),b_n^{(m)}(j;1)) \times \cdots \times [a_n^{(m)}(j;d),b_n^{(m)}(j;d)) \subseteq \mathcal{X}$. A similar relation holds for the sequential query strategy: replace $n$ by $n_t$, and the player index $m$ by the player being queried at sub-instant $n_t$.
\end{assumption}
When the query region $A_n^{(m)}$ is written this way, the space in which $A_n^{(m)}$ lies in is identified with the space $\mathcal{A}_n^{(m)}=\{(a(j;l),b(j;l)):j=1,\dots,J_n^{(m)}, l=1,\dots,d, a(j;l)\leq b(j;l)\}$, a closed subset of $\RR^{2 d J_n^{(m)}}$. Then, for a fixed initial distribution $p_0(\cdot)$, the posterior distribution $p_n(\cdot)$ can be identified with the set $\{(a_k^{(m)}(j;l),b_k^{(m)}(j;l)),Y_{k+1}^{(m)}:m=1,\dots,M,k=0,\dots,n-1\}$ that lies in the space $\mathcal{S}_n=(\mathcal{A}_0 \times \cdots \times \mathcal{A}_{n-1}) \times \mathcal{Y}^n$. It follows that the mapping from $p_n \in \mathcal{S}_n$ to $H(p_n)\in \RR$ is measurable. A similar set of assumptions were made in \cite{Jedynak12} to guarantee measurability. Another possible parametrization of the query regions $A_n^{(m)}$ that makes the results in this paper valid is half-spaces, i.e., $A_n^{(m)}=\{x\in \mathcal{X}: (a_{n}^{(m)})^T x \leq b_n^{(m)}\}$.

Define the set of dyadic partitions of $\RR^d$, induced by the queries $\{A^{(m)}\}_m$:
\begin{equation} \label{eq:dyadic_partition}
	\gamma(A^{(1)},\dots,A^{(M)}) = \left\{ \bigcap_{m=1}^M (A^{(m)})^{i_m}: i_m \in \{0,1\} \right\}
\end{equation}
where $(A)^0:=A^c$ and $(A)^1:=A$. The cardinality of this set of subsets is $2^M$ and each of these subsets partition $\RR^d$. 

Define the density parametrized by $\bA_n,\mathcal{F}_n,i_1,\dots,i_M$, for the joint query strategy:
\begin{equation*}
	g_{i_1:i_M}(y^{(1)},\dots,y^{(M)}|\bA_n,\mathcal{F}_n):= \prod_{m=1}^M f_{i_m}^{(m)}(y^{(m)}|A_n^{(m)},\mathcal{F}_n)
\end{equation*}
where $i_j\in \{0,1\}$.

\subsection{Equivalence Theorem}

We first establish the structure of the optimal joint policy using tools from stochastic control theory. The proof is based on Bellman's optimality principle.
\begin{theorem} (Joint Optimality Conditions, Known Error Probabilities) \label{thm:joint}
	Under Assumption \ref{assump:cond_indep}, an optimal joint policy that minimizes the Shannon entropy of the posterior distribution $p_n$ achieves the following entropy loss:
	\begin{align}
		G^* &= \sup_{A^{(1)},\dots,A^{(M)}} \Big\{ H\left( \sum_{i_1:i_M=0}^1 g_{i_1:i_M}(\cdot) P_n\Big(\bigcap_{m=1}^M (A_n^{(m)})^{i_m}\Big) \right) \nonumber \\
		 &- \sum_{i_1:i_M=0}^1 H\left(g_{i_1:i_M}(\cdot)\right) P_n\Big(\bigcap_{m=1}^M (A_n^{(m)})^{i_m}\Big) \Big\},         \label{eq:joint_optimal}
	\end{align}
	where $H(f)$ is the Shannon entropy of the probability mass function $f$.
\end{theorem}

Theorem \ref{thm:joint} generalizes the bisection policy \cite{Jedynak12, CastroNowak07} to multiple players. The fusion rule is a posterior update and by Bayes rule, we have:
\begin{equation} \label{eq:Bayes}
	p_{n+1}(x) \propto \PP(\bY_{n+1}=\by_{n+1}|\bA_n,X^*=x,\mathcal{F}_n) \times  p_n(x)
\end{equation}
where $\by_{n+1} \in \{0,1\}^M$ are the $M$ observations at time $n$. Next we establish that a sequential query strategy achieves the same average entropy reduction as that of the optimal joint query strategy.

\begin{figure}[ht]
	\centering
		\includegraphics[width=0.50\textwidth]{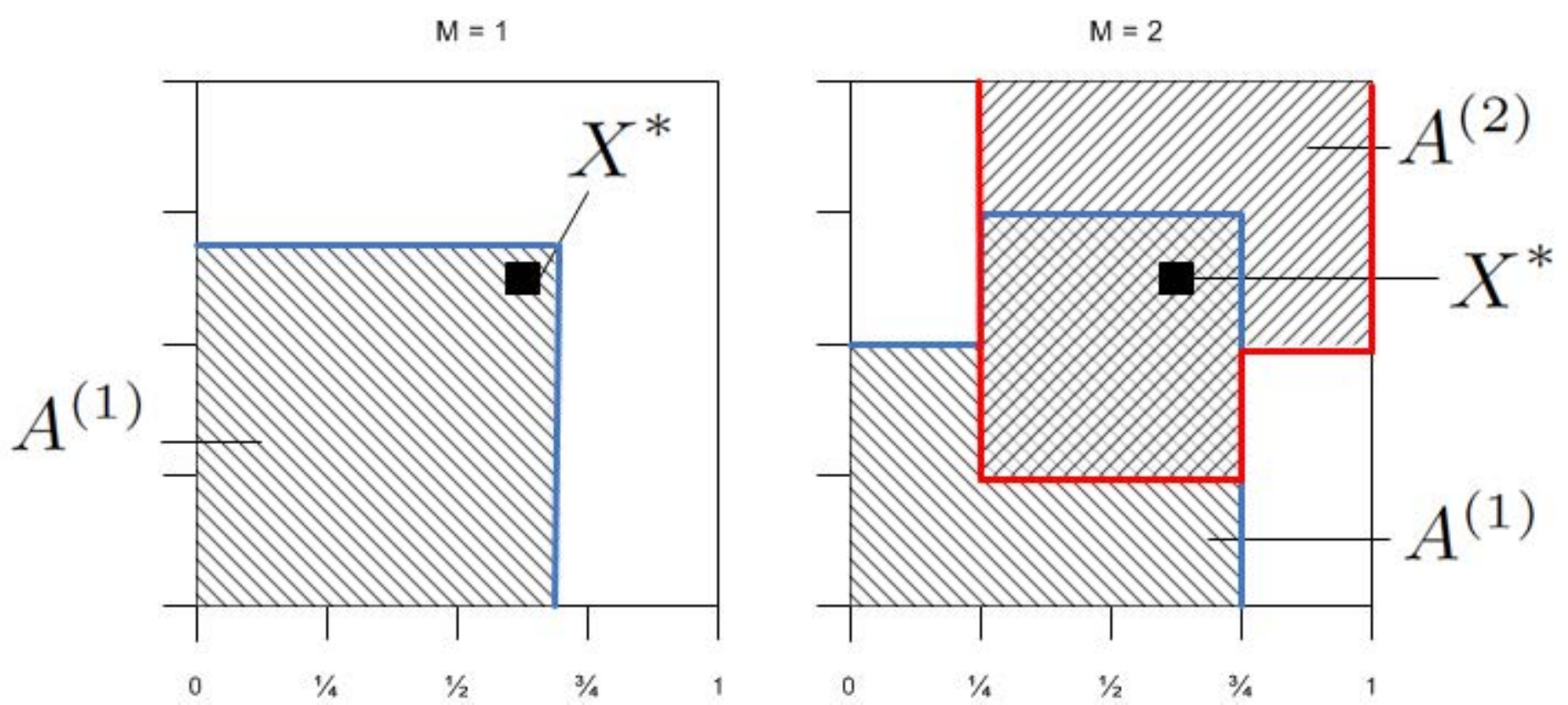}
	\caption{\small{Jointly optimal queries under a uniform prior for two dimensional target search. The target $X^*$ is indicated by a black square. The one-player bisection rule (left) satisfies the optimality condition (\ref{eq:optimal_joint_law}) with optimal query $A^{(1)}=[0,\frac{1}{\sqrt{2}}]\times [0,\frac{1}{\sqrt{2}}]$. The two-player bisection rule (right) satisfies (\ref{eq:optimal_joint_law}) with optimal queries $A^{(1)}=[0,\frac{3}{4}]\times[0,\frac{1}{2}] \cup [\frac{1}{4},\frac{3}{4}]\times[\frac{1}{2},\frac{3}{4}], A^{(2)}=[\frac{1}{4},1]\times[\frac{1}{2},1]\cup [\frac{1}{4},\frac{3}{4}]\times[\frac{1}{4},\frac{1}{2}]$. We note that using the policy on the left, if player 1 responds that $X^* \in [0,\frac{1}{\sqrt{2}}]\times [0,\frac{1}{\sqrt{2}}]$, with high probability, then the posterior will concentrate on that region. When using the policy on the right, if player 1 and 2 respond that $X^*\in A^{(1)} \cap A^{(2)}$ with high probability, then the posterior will concentrate more on the intersection of the queries, thus better localizing the target as compared with the single player policy. } }
	\label{fig:optimal_query_2d}
\end{figure}

Next, we prove the equivalence theorem that shows the maximal entropy loss of the joint query design is the same as the entropy loss of sequential query design. This is one of the principal results of the paper.
\begin{theorem} (Equivalence, Known Error Probabilities) \label{thm:separation}
	Under Assumptions \ref{assump:cond_indep}, \ref{assump:BSC} and \ref{assump:query_region_regularity}:
\begin{enumerate}
	\item The expected entropy loss under an optimal joint query design is the same as the sequential query design. This loss is given by:
		\begin{equation} \label{eq:entropy_loss}
			C = \sum_{m=1}^M C(\epsilon_m) = \sum_{m=1}^M (1-h_b(\epsilon_m)),
		\end{equation}
		where $h_b(\epsilon_m)=-\epsilon_m \log(\epsilon_m) - (1-\epsilon_m) \log(1-\epsilon_m)$ is the binary entropy function.
	\item All jointly optimal control laws equalize the posterior probability over the dyadic partitions induced by $\bA_n=\{A_n^{(1)}, \ldots, A_n^{(M)}\}$:
		\begin{equation} \label{eq:optimal_joint_law}
			P_n(R) = \int_{R} p_n(x) dx = 2^{-M}, \forall R \in \gamma(\bA_n),
		\end{equation}
		where the set $\gamma(\cdot)$ was defined in (\ref{eq:dyadic_partition}).
\end{enumerate}
\end{theorem}

Theorem \ref{thm:separation} shows that the optimal joint policy can be determined and implemented using the simpler greedy sequential query design.  Note that, despite the fact that all players are conditionally independent, the joint policy does not decouple into separate single-player optimal policies. This is analogous to the non-separability of the optimal vector-quantizer in source coding even for independent sources \cite{Gray}. In addition, the optimal queries must be overlapping-i.e., $\bigcap_{m=1}^M A_n^{(m)} \neq \emptyset$, but not identical. Finally, we remark that the optimal query $\bA_n$ is not unique, so it is possible that there exists an even simpler optimal control law than the sequential greedy policy.

We note that, considering the sequential query design, if the fusion center had the choice of asking one player at each time, then the optimal selection scheme would be to choose the player with the minimum BSC crossover probability. This equivalence theorem is a stepping stone to the unknown noisy channel case where it is not clear which player is most accurate.

We finally remark that the equivalence theorem also holds for non-symmetric binary-output channels, with appropriate modifications in the non-dyadic partition structure of the optimality conditions (\ref{eq:optimal_joint_law}). For simplicity, the rest of the paper focuses on the binary symmetric channel (BSC).


\subsubsection{Equivalence: Intuition}
A simple intuitive way to see the equivalence property stated in Theorem \ref{thm:separation} is through the chain rule of the mutual information. Consider the joint query strategy and its associated filtration $\mathcal{F}_n$. According to Theorem \ref{thm:joint}, the optimal policy is to choose the queries such that the conditional mutual information is maximized. The chain rule of conditional mutual information \cite{CoverThomas} implies:
\begin{align*}
	I(X^*;& \bY_{n+1}|\bA_n,\mathcal{F}_n) = I(X^*; Y_{n+1}^{(1)}|A_n^{(1)},\mathcal{F}_n) \\
	&+ \sum_{m=2}^{M} I(X^*; Y_{n+1}^{(m)}|A_n^{(m)},\{A_n^{(k)},Y_{n+1}^{(k)}\}_{k=1}^{m-1},\mathcal{F}_n),
\end{align*}
which relates the joint mutual information of the LHS (as in the joint scheme) to the mutual information of each player conditioned on the responses of the previous players (as in the sequential scheme). Letting $M=2$ for concreteness, we observe:
\begin{align*}
	&I(X^*; Y_{n+1}^{(1)}, Y_{n+1}^{(2)}|A_n^{(1)},A_n^{(2)},\mathcal{F}_n) \\
		&= I(X^*;Y_{n+1}^{(2)}|Y_{n+1}^{(1)},A_n^{(2)},A_n^{(1)},\mathcal{F}_n) + I(X^*;Y_{n+1}^{(1)}|A_n^{(1)},\mathcal{F}_n),
\end{align*}
This relation implies that the mutual information between the target $X^*$ and the response $Y_{n+1}^{(2)}$ of the second player depends on the response of the first player $Y_{n+1}^{(1)}$. It follows that the information available for query design $A_n^{(2)}$ for the second player is larger than the information available for query design $A_n^{(1)}$ for the first player.

\subsubsection{Equivalence: One-dimensional Example}

\begin{figure}[ht]
	\centering
		\includegraphics[width=0.35\textwidth]{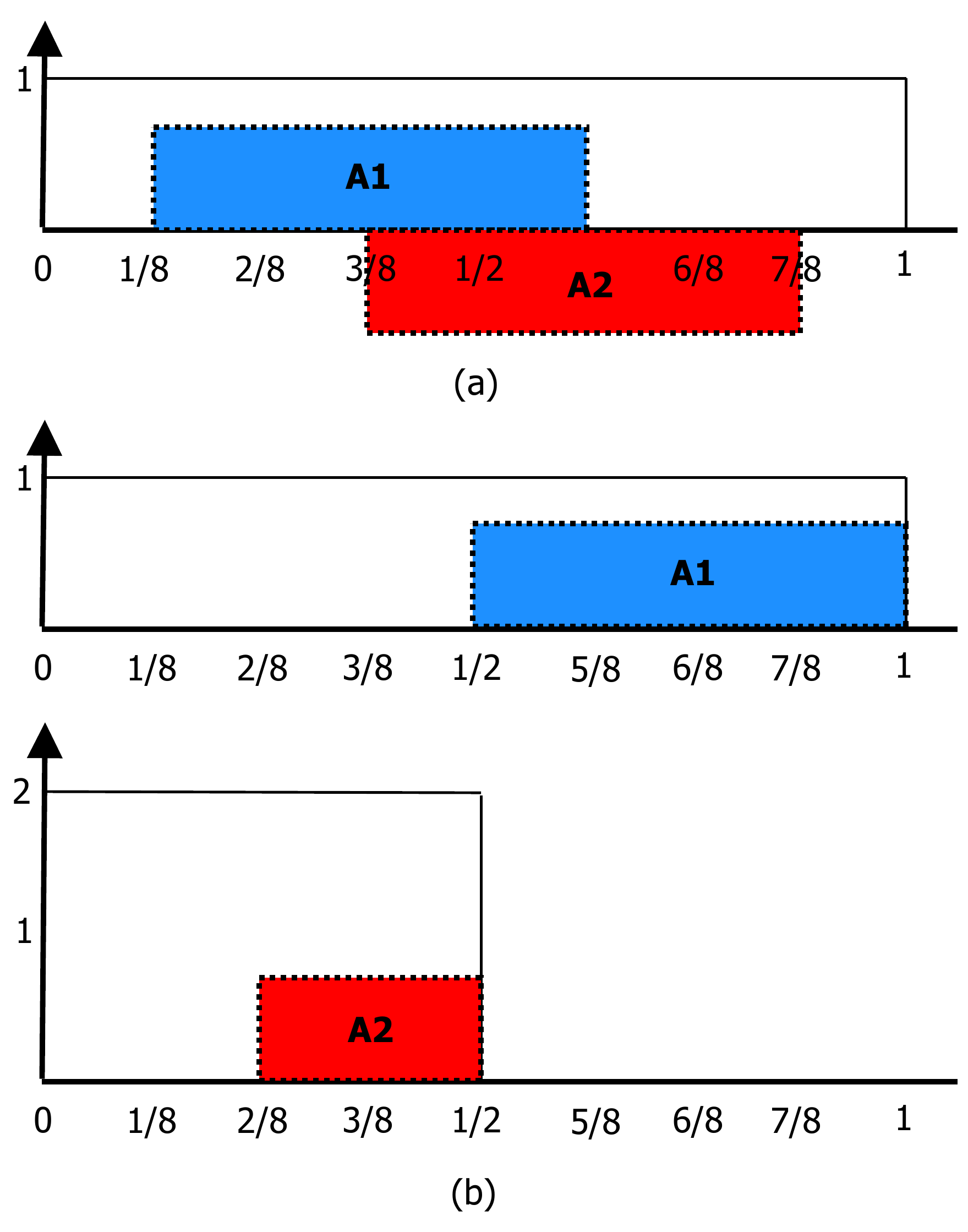}
	\caption{\small{ Illustration of jointly optimal policy (a) and sequential policy (b) for one-dimensional target, uniformly distributed over $[0,1]$, and two players. In each case the total length of the intervals not covered by the queries (uncertainty) is equal to $1/4$.  }}
	\label{fig:illustration_d1_M2}
\end{figure}

As a specific example, let us consider the one-dimensional case with $M=2$ collaborating players. Consider the query design problem for this case. We assume that the prior density $p_0$ is uniform over the position of a target in one dimension, i.e., the target state is in the domain $\mathcal{X}=[0,1]$. We define the queries as intervals-i.e., $A_n^{(1)}=[a,b]$ and $A_n^{(2)}=[c,d]$. The optimal policy (\ref{eq:optimal_joint_law}) requires the queries to be overlapping and so we impose the constraints $a<c, c<b$ and $b<d$. Choosing $a=1/8, b=1/2+1/8, c=1/2-1/8$ and $d=1-1/8$, we observe that the optimality conditions in (\ref{eq:optimal_joint_law}) are satisfied over the dyadic partition set $\gamma(\mathbf A_n)=\{A_n^{(1)} \cap A_n^{(2)}$, $A_n^{(1)} \cap \overline{A}_n^{(2)}$, $A_n^{(1)} \cap \overline{A}_n^{(2)}$ and $\overline{A}_n^{(1)} \cap \overline{A}_n^{(2)}\}$. Thus, this is a jointly optimal law and is illustrated graphically in Figure \ref{fig:illustration_d1_M2} (a). We note that the region of uncertainty has size $1/4$ (region not covered by queries).

The sequential policy consists of a sequence of bisections. This policy is illustrated in Figure \ref{fig:illustration_d1_M2} (b) and the region of uncertainty also has size $1/4$.

\subsection{Costs \& Player Selection}

In this subsection, we consider the minimum expected entropy criterion with costs. At each time instant, we are allowed to choose one out of the total $M$ players to query, and for each player being queried, a cost is incurred \footnote{This setup can be trivially extended to choosing $M_0$ players out of $M$.}. This problem formulation trades off the reduction in uncertainty on the target's location with the cost for obtaining information.

In the context of the `machine+human' problem, the addition of a cost function for using each sensor allows us to incorporate human fatigue factors into the design of the algorithm as well as the cost of using the machine. In the context of agile sensing, a subset of the most informative sensors needs to be selected for query design, while taking into account the cost of each sensor. Often, higher accuracy sensors are more costly to use.

A similar problem has been recently studied by Sznitman, et al. \cite{Sznitman:2013}. Our formulation differs from the work of \cite{Sznitman:2013} as we consider the joint problem of selecting the player to query and the optimal query associated with him, in addition to having query-dependent costs.

Define the sum of costs incurred after $n$ iterations:
\begin{equation*}
	T_n = \sum_{k=0}^{n-1} K_k(u_k,A_k),
\end{equation*}
where $K(u,A)$ is the cost associated with the $u$th player and $A$ is the associated query. 
We propose to find optimal collaborative query policies $\zeta$, that account for player costs, by solving:
\begin{equation} \label{eq:min_entropy_cost}
	\inf_{\zeta} \EE^{\zeta} \left[H(p_N) + \gamma T_N\right],
\end{equation}
where $\gamma\geq 0$ controls the cost-performance tradeoff. Here $\zeta$ is a control policy whose actions include selection of the active player at time $k$, i.e., $u_k$, and the associated query $A_k$.
Define the value function $V_n(p_n,T_n)$ as:
\begin{equation} \label{eq:value_def}
	V_n(p,t) = \inf_{\zeta} \EE^\zeta[H(p_N)+\gamma T_N|p_n=p,T_n=t].
\end{equation}
Policies that achieve this value are optimal. The Bellman optimality principle \cite{Dynkin:1979} implies that the solution to (\ref{eq:value_def}) can be found by backwards induction:
\begin{equation}
	V_n(p,t) = \inf_{u,A} \EE[ V_{n+1}(p_{n+1},T_{n+1}) | p_n=p, T_n=t, u,A ]. \label{eq:value_recursion}
\end{equation}
In particular, a policy that attains the infimum in (\ref{eq:value_recursion}) for all $n,p,t$ also attains the infimum in (\ref{eq:min_entropy_cost}) \cite{Dynkin:1979}.
%
%
Define the gain function:
\begin{equation*}
	G_k(u,A) = I(X^*;\bY_{k+1}|u_k=u,A_k=A,p_k) - \gamma K_k(u,A).
\end{equation*}
The next theorem characterizes the structure of optimal policies and the entropy loss at each stage.
\begin{theorem} \label{thm:policy_cost}
	All optimal control laws under the criterion (\ref{eq:min_entropy_cost}) achieve the supremum:
	\begin{equation} \label{eq:optim_condition}
		\sup_{1\leq u\leq M } \sup_{A} G_k(u,A).
	\end{equation}
	The value function for the problem (\ref{eq:min_entropy_cost}) is:
	\begin{align}
		V_N(p_N,T_N) &= H(p_N) + \gamma T_N \nonumber \\
		V_n(p_n,T_n) &= H(p_n) + \gamma T_n - \sum_{k=n}^{N-1} \sup_{u,A} G_k(u,A) \nonumber \\
			&\qquad 0\leq n\leq N-1 \label{eq:value_function_cost}
	\end{align}
\end{theorem}

The solution can be simplified if the cost does not depend on the query.

\begin{corollary} \label{cor:min_entropy_cost}
	Consider the problem (\ref{eq:min_entropy_cost}) and assume that the cost is independent of the query, i.e., $K_k(u,A)=K_k(u)$. Then the control policy is optimal if, for $n=1, \ldots, N$, the players are selected according to: 
	\begin{equation} \label{eq:balance}
		\max_{u\in \{1,\dots,M\}} \left\{C_n(u) - \gamma K_n(u) \right\},
	\end{equation}
	where $C_n(u)$ is the capacity of the $u$th player and the associated query satisfies the condition of Theorem \ref{thm:separation} (i.e., is a bisection of the posterior density).
\end{corollary}
Corollary \ref{cor:min_entropy_cost} makes the tradeoff between entropy loss and cost apparent. This scenario is relevant in the setting where high quality sensors might be too costly to use, while less informative sensors might be cheaper. The criterion (\ref{eq:balance}) provides a way to balance this tradeoff through the parameter $\gamma>0$. 

\section{Mean-Square Error Performance Bounds} \label{sec: bound_MSE}
In this section, we provide exponential lower and upper bounds on the MSE of the sequential Bayesian estimator in Section \ref{sec: multiplayer_entropy_loss}.D.

\subsection{Lower Bounds via Entropy Loss}
Theorem \ref{thm:separation} yields the value of the cooperative game in terms of expected entropy reduction, which is the sum of the ``capacities'' \footnote{The ``capacity'' of each player is the Shannon channel capacity of each BSC \cite{CoverThomas}.} of all the players. This value function is used next to provide a lower bound on the MSE of the sequential Bayesian estimator.

\begin{theorem} (Lower Bound on MSE) \label{thm:lb}
	Let Assumptions \ref{assump:cond_indep}, \ref{assump:BSC} and \ref{assump:query_region_regularity} hold. Assume the entropy $H(p_0)$ is finite. Then, the MSE of the joint or sequential query policies in Theorems 1 and 2 satisfies:
	\begin{equation} \label{eq:lb}
		\frac{K}{2\pi e} d \exp\left(-\frac{2nC}{d}\right)  \leq \EE[\parallel X^*-X_n\parallel_2^2]
	\end{equation}
	where $K=e^{2H(p_0)}$, $d$ is the dimension of the target space and $X_n$ is the median of the posterior distribution $p_n(\cdot)$. The expected entropy loss per iteration is $C=\sum_m C(\epsilon_m)$.
\end{theorem}
Observe that the bound in (\ref{eq:lb}) holds for any policy $\zeta$, and for optimal policies $\zeta^*$ the bound becomes tighter since $\EE^\zeta[H(p_n)]=H(p_0)-nC$ for this case. We also note that the bound behaves exponentially as a function of the number of queries $n$ with rate exponent given by the sum of the capacities $C$.

\subsection{Upper Bounds}

The performance analysis of the probabilistic bisection algorithm (PBA) is difficult primarily due to the continuous nature of the posterior \cite{CastroNowak07}. A discretized version of PBA was first proposed in \cite{BZ}, known as the Burnashev-Zingagirov (BZ) algorithm, which imposes a piecewise constant structure on the posterior distribution. The BZ algorithm assumes one-dimensional targets and requires an initial distribution to begin and a query-response mechanism. Due to the discretization of the posterior distribution, the one-dimensional query will either overestimate or underestimate the median of the true continuous posterior distribution. Thus, a biased coin flip (where the bias depends on the posterior distribution) is required at each iteration to choose the query point. For more details on the BZ algorithm, the interested reader can refer to \cite{CastroNowak07} and Appendix A in \cite{CastroPhD}. 

For simplicity of discussion, we assume the target location is constrained to the unit interval $\mathcal{X}=[0,1]$. The generalization to $d>1$ is a challenging open problem. A step size $\Delta>0$ is defined such that $\Delta^{-1}\in \NN$ and the posterior after $j$ iterations is $p_j:\mathcal{X}\to \RR$, given by
\begin{equation*}
	p_j(x) = \frac{1}{\Delta} \sum_{i=1}^{\Delta^{-1}} a_i(j) I(x\in I_i)
\end{equation*}
where $I_1 = [0,\Delta], I_i=((i-1)\Delta,i \Delta]$ for $i=2,\dots,\Delta^{-1}$. We define the discretized posterior at time $j$ as the probability vector $\ba(j)=[a_1(j),\dots,a_{\Delta^{-1}}(j)]$. The initial posterior is $a_i(0)=\Delta, \forall i$. The posterior is characterized completely by the discretized posterior $\ba(j)$ which is updated at each iteration via Bayes rule \cite{CastroPhD}.

Convergence rates were derived for the one-dimensional case in \cite{CastroNowak07} for the bounded noise case (i.e., constant error probability) and for the unbounded noise case (i.e., error probability depends on distance from target $X^*$ and converges to $1/2$ as the estimate reaches the target) in \cite{CastroNowak06}. A modified version of this algorithm that is proven to handle unbounded noise was shown in \cite{CastroNowak06}. Theorem \ref{thm:ub} derives upper bounds on MSE using ideas from \cite{CastroNowak06}.

First, we need a simple lemma.
\begin{lemma} \label{lemma: expectation_bound}
	Let $\hat{X}_n$ be an estimator of target $X^*$ lying in domain $[0,1]$. Then, for all $\Delta \in [0,1]$, we have:
	\begin{equation*}
		\EE[(X^*-\hat{X}_n)^2] \leq \Delta^2 + (1-\Delta^2) \PP(|X^*-\hat{X}_n|>\Delta)
	\end{equation*}
\end{lemma}

Now, we are in a position to prove the upper bound on the MSE using Lemma \ref{lemma: expectation_bound}.
\begin{theorem} (Upper Bound on MSE) \label{thm:ub}
	Consider the sequential bisection algorithm for $M$ players in one-dimension, where each bisection is implemented using the BZ algorithm. Then, we have:
	\begin{align}
		\PP(|X^*-\hat{X}_n|>\Delta) \leq (\frac{1}{\Delta}-1) \exp\left(-n \bar{C} \right) \nonumber \\
		\EE[(X^*-\hat{X}_n)^2] \leq (2^{-2/3}+2^{1/3}) \exp\left(-\frac{2}{3} n \bar{C}\right) \label{eq:ub}
	\end{align}
	where $\bar{C}=\sum_{m=1}^M \bar{C}(\epsilon_m)$, $\bar{C}(\epsilon)=1/2 - \sqrt{\epsilon (1-\epsilon)}$.
\end{theorem}
The combination of the lower bound (Theorem \ref{thm:lb}) and the upper bound (Theorem \ref{thm:ub}) imply that the MSE goes to zero at an exponential rate with rate constant between $2C$ and $\frac{2}{3} \bar{C}$.

\section{Application: Human-in-the-loop} \label{sec: human_in_the_loop}
In this section, we apply our methodology to a simple human-aided sensing problem. We first consider the case of two players, one human and a one machine, and then generalize to the case of an arbitrary number of humans and machines. As a concrete example, consider the problem of chemometric toxin-detection. A machine (robot) and a human hazmat expert are sent to a remote location where there has been a release of some unknown toxin, e.g., a bio-toxin released in a gas leak or fluid spill. On-site mass spectroscopy yields an energy spectrum that can be used to identify the bio-toxin by detecting locations of spectral peaks. The abilities of the robot and the human are complementary for this peak localization task: the robot can quantify a peak in the spectrum with very fine spectral resolution but cannot easily distinguish between true and false peaks, which are more easily disambiguated by the human chemometrics expert. Our centralized 20 questions controller asks the robot and human an increasingly refined sequence of binary questions about the location of the bio-toxin peak in the measured spectrum. The errors in the responses to these questions are binary and error prone, where the different probability of errors of the human and robot reflect their relative strengths and weaknesses.

\subsection{Case of Two Players}

\begin{figure}[ht]
	\centering
		\includegraphics[width=0.50\textwidth]{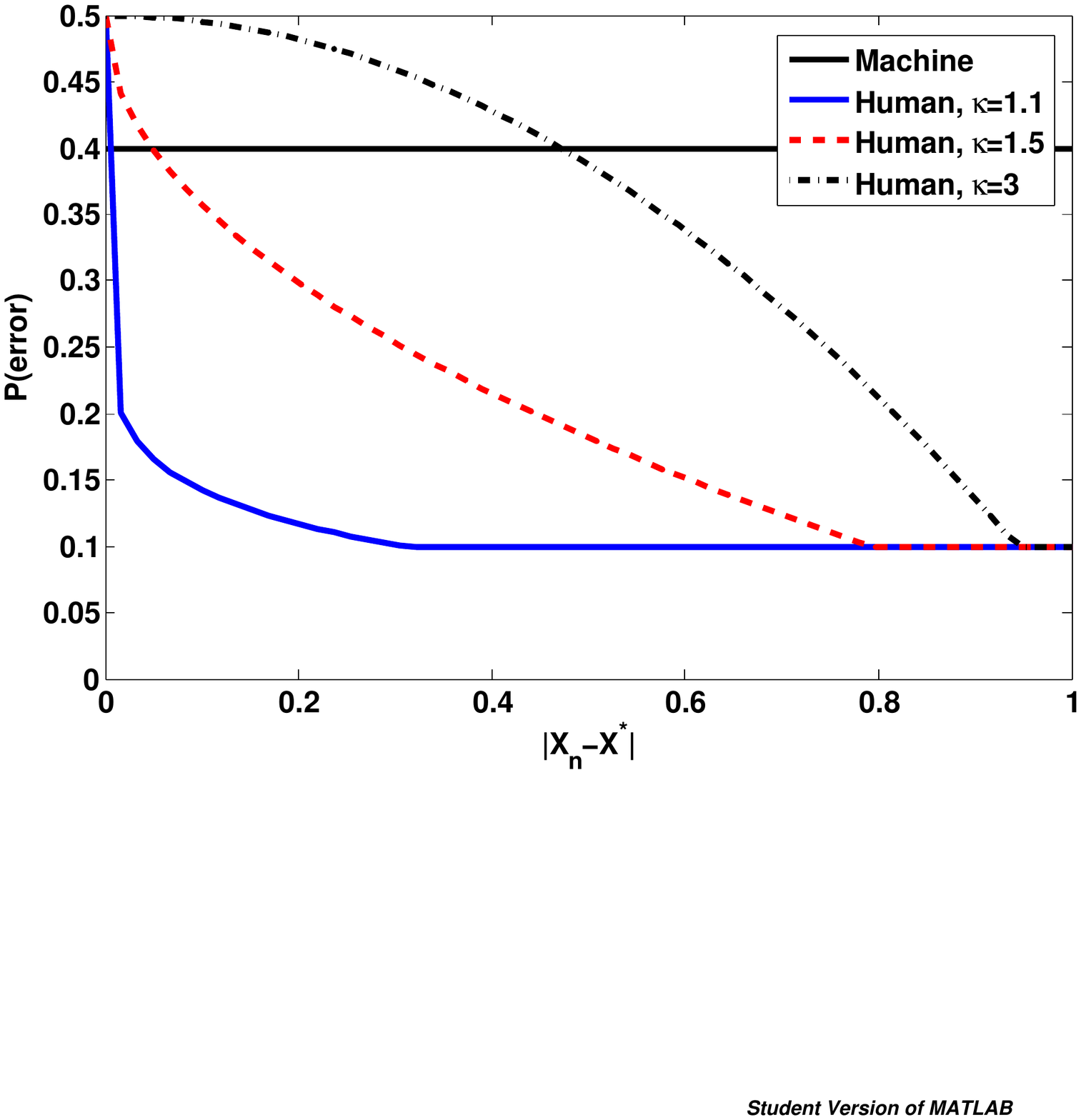}
	\caption{\small{Human error probability as a function of distance from target $|X^*-X_n|$ for $\delta_0=0.4, \mu=0.45$ and various $\kappa>1$.}}
	\label{fig:human_error_model}
\end{figure}

We first consider the two-player case where player 1 (the machine) has a constant error probability $\epsilon_1\in (0,1/2)$ and player 2 (the human) has error probability increasing as the target localization error decreases:
\begin{equation} \label{eq:human_err}
	\PP(Y_{n+1}^{(2)}\neq z|Z_n^{(2)}=z) = \frac{1}{2} - \min(\delta_0,\mu |X^*-X_n|^{\kappa-1})
\end{equation}
where $\kappa>1, 0<\delta_0<\mu<1/2$ are reliability parameters characterizing the human ability to localize the target $X^*$\footnote{The parameter $\kappa$ controls the resolution of the human. It becomes increasingly difficult for the human to decide between close hypotheses as $\kappa$ goes to infinity.}. Figure \ref{fig:human_error_model} illustrates the human error model as a function of $|X^*-X_n|$. This is a popular model used for human-based optimization \cite{Jamieson12} and active learning of threshold functions \cite{CastroNowak06}. From the nature of the error probability (\ref{eq:human_err}), and under the coarse-to-fine sequential bisection policy in Section \ref{sec: bound_MSE}, we expect that the answers provided by the human will be helpful in the beginning iterations but their value will go to zero as the number of iterations grows to infinity. This is because the human propensity for error becomes larger as the questions become more refined and location of the target more difficult to resolve with precision.

In the context of the bio-toxin detection example, model (\ref{eq:human_err}) reflects the decreasing ability of the human to answer increasingly precise questions about the location of the bio-toxin spectral peak. In particular, the parameter $\delta_0$ in (\ref{eq:human_err}) is the error floor for the human, which satisfies $\epsilon_1>1/2-\delta_0$ to represent the coarse resolution advantage of the human as compared to the machine. The scale and shape parameters $\mu$ and $\kappa$ specify the spectral resolution threshold $|X^*-X_n| = (\delta_0/\mu)^{1/{\kappa-1}}$ below which the human starts to lose ground to the machine.

Using a similar argument as in the proof of Theorem \ref{thm:ub}, and using the modified BZ algorithm \cite{CastroNowak06}, from Lemma 1 in \cite{CastroNowak06}, we have the following. For $\kappa\geq 2$ with $\alpha_1 = \frac{\sqrt{\epsilon_1}}{\sqrt{\epsilon_1}+\sqrt{1-\epsilon_1}}, \alpha_2=0.09\mu (3\Delta/4)^{\kappa-1}$:
\begin{equation*}
	\PP(|X^*-\hat{X}_n|>\Delta) \leq \Delta^{-1} \exp\left(-n \left[\bar{C}(\epsilon_1) + \frac{\mu^2}{50} \left(\frac{3\Delta}{4}\right)^{2\kappa-2}\right] \right).
\end{equation*}
Applying our Lemma \ref{lemma: expectation_bound} in Section \ref{sec: bound_MSE}, this leads to the MSE upper bound dependent on $\Delta$:
\begin{align} 
	\EE&[(X^*-\hat{X}_n)^2] \leq \Delta^2 \nonumber \\
	&+ \Delta^{-1} \exp\left(-n \left[\bar{C}(\epsilon_1) + \frac{\mu^2}{50} \left(\frac{3\Delta}{4}\right)^{2\kappa-2}\right] \right) \label{eq:p1_human_bnd}
\end{align}
With the choice $\Delta=2^{-1/3} e^{-n\bar{C}(\epsilon_1)/3}$,
\begin{align}
 & \EE[(X^*-\hat{X}_n)^2] \leq \exp\left( -\frac{2}{3}n\bar{C}(\epsilon_1) \right) \nonumber \\
 	& \times \left[ 2^{-2/3}+2^{1/3}\exp\left(-\frac{\mu^2}{50} \Big( \frac{3\cdot 2^{-1/3}}{4}\Big)^{2\kappa-2} n e^{-n\bar{C}(\epsilon_1)\frac{2\kappa-2}{3}} \right) \right] \label{eq:ub_human}
\end{align}
which is no greater than the machine alone MSE bound (compare (\ref{eq:ub_human}) with (\ref{eq:ub})). Asymptotically as $n\to\infty$, the two bounds both converge to zero at the same rate.

\begin{figure}[ht]
	\centering
		\includegraphics[width=0.50\textwidth]{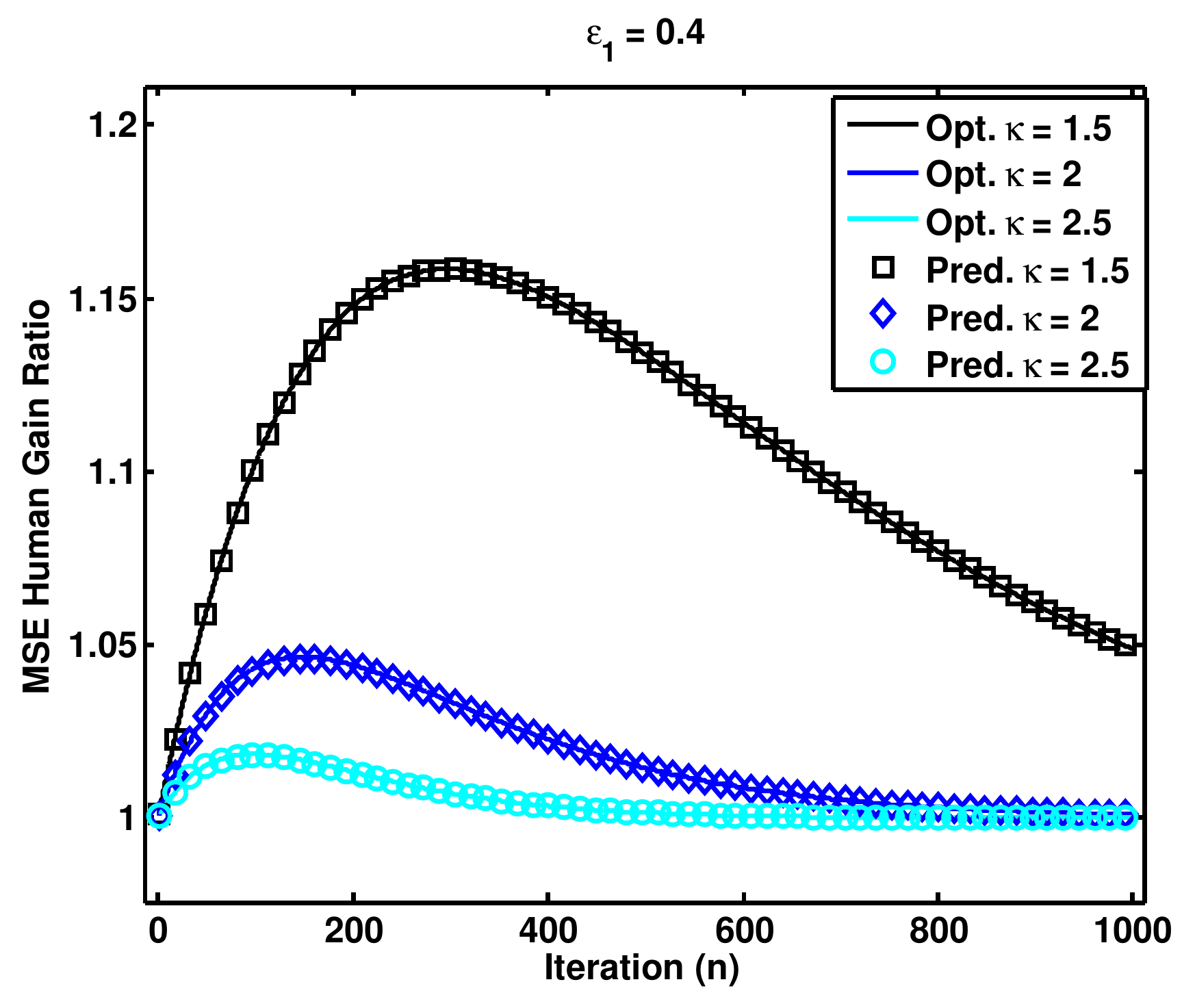}
	\caption{Human gain ratio (see Eq. (\ref{eq:hgratio})) for a pair of players consisting of a human and machine. The human provides the largest gain in the beginning few iterations and the value of information decreases as $n\to\infty$. The circles are the predicted curves according to (\ref{eq:ub_human}), while the solid lines are the optimized versions of the bound (\ref{eq:p1_human_bnd}) (as a fuction of $\Delta$) for each $n$. The predictions well match the optimized bounds.}
	\label{fig:p1hum_kappa}
\end{figure}

We define the human gain ratio (HGR) as the ratio of MSE upper bounds associated with machine and machine+human, respectively, given by
\begin{equation} \label{eq:hgratio}
	R_n(\kappa) = \frac{2^{-\frac{2}{3}}+2^{\frac{1}{3}}}{2^{-\frac{2}{3}}+2^{\frac{1}{3}}\exp\left(-\frac{\mu^2}{50} (\frac{3\cdot 2^{-1/3}}{4})^{2\kappa-2} n e^{-n\bar{C}(\epsilon_1)\frac{2\kappa-2}{3}}\right) }
\end{equation}
The HGR is plotted in Figure \ref{fig:p1hum_kappa} as a function of $\kappa$. This analysis quantifies the value of including the human-in-the-loop for a sequential target localization task. We note that the larger $\epsilon_1$ is, the larger is the HGR. Also, as $\kappa$ decreases to 1, the ratio increases, meaning that the human accuracy approaches that of the machine.

\subsection{Case of More than Two Players}

This result is generalized in the next corollary to the case of multiple machines and multiple humans. 
Consider the sequential bisection algorithm for $M=M_1+M_2$ players in one-dimension, where each bisection is implemented using the BZ algorithm. Here, there are $M_1$ machines with constant error probabilities $\{\epsilon_m\}$ and $M_2$ humans with non-constant error probabilities parameterized by parameters $\{\mu_l,\delta_{0,l},\kappa_l\}$. Then, we have:
	\begin{align}
		\PP&(|X^*-\hat{X}_n|>\Delta) \leq \left(\frac{1}{\Delta}-1\right) \nonumber \\
			&\times \exp\left(-n \left\{ \bar{C} + \sum_{l=1}^{M_2} \frac{\mu_l^2}{50} \left(\frac{3\Delta}{4}\right)^{2\kappa_l-2} \right\} \right) \nonumber \\
		\EE&[(X^*-\hat{X}_n)^2] \leq \exp\left(-\frac{2}{3}n\bar{C}\right) \nonumber \\
			&\times \left[2^{-\frac{2}{3}} + 2^{\frac{1}{3}} \exp\left(-\sum_{l=1}^{M_2} \frac{\mu_l^2}{50} \left(\frac{3 \cdot 2^{-\frac{1}{3}}}{4}\right)^{2\kappa_l-2} n e^{-n \bar{C}\frac{(2\kappa_l-2)}{3}} \right) \right] \nonumber
	\end{align}
	where $\bar{C}=\sum_{m=1}^{M_1} \bar{C}(\epsilon_m)$, $\bar{C}(\epsilon)=1/2 - \sqrt{\epsilon (1-\epsilon)}$.

\section{Noisy 20 Questions with Collaborative Players: Unknown Error Probability} \label{sec: unknown_error_probability}

In this section we consider the setting where the error probabilities $\{\epsilon_m\}_{m=1}^M$ of the $M$ players are unknown. In this case, the Bayes posterior update $p_n \mapsto p_{n+1}$ required for implementing the optimal policy in Sec. \ref{sec: multiplayer_entropy_loss} is not implementable as it requires knowledge of the error probabilities. Here we consider the alternative when the unknown $\epsilon_m \in (0,1/2)$ and uniformly distributed. For this case we propose a joint estimation scheme to estimate the target $X^*$ and the error probabilities $\bepsilon^*=(\epsilon_1^*,\dots,\epsilon_M^*)$. The method propagates the joint posterior distribution of the joint random vector $(X^*,\bepsilon^*)$ forward in time given the designed queries and noisy responses. 

Define the random vector $\bepsilon=(\epsilon_1,\dots,\epsilon_M) \in [0,1/2)^M$ and the joint posterior distribution $\PP(X^*=x,\bepsilon^*=\bepsilon|\mathcal{F}_n)=p_n(x,\bepsilon)$. We consider policies that minimize the expected entropy (\ref{eq: min_entr_obj}).

\subsection{Assumptions}
We make an analogous conditional independence assumption to Assumption \ref{assump:cond_indep} for the unknown channel case.
\begin{assumption} \label{assump:cond_indep_eps}
We assume that the players' responses are conditionally independent:
\begin{align*}
	\PP(&\bY_{n+1}=\by|\bA_n,X^*=x,\bepsilon^*=\bepsilon,\mathcal{F}_n) \\
		&= \prod_{m=1}^M \PP(Y_{n+1}^{(m)}=y^{(m)}|A_n^{(m)},X^*=x,\epsilon_m^*=\epsilon_m,\mathcal{F}_n),
\end{align*}
where
\begin{align*}
	\PP(&Y_{n+1}^{(m)}=y^{(m)}|A_n^{(m)},X^*=x,\epsilon_m^*=\epsilon_m,\mathcal{F}_n) \\
		&= \Bigg\{ \begin{array}{ll} f_1^{(m)}(y^{(m)}|\epsilon_m,A_n^{(m)},\mathcal{F}_n), & x\in A_n^{(m)} \\ f_0^{(m)}(y^{(m)}|\epsilon_m,A_n^{(m)},\mathcal{F}_n), & x\notin A_n^{(m)} \end{array}.
\end{align*}
\end{assumption}

\subsection{Sequential Query Design}
As in the case of known $\{\epsilon_m\}$, in the sequential setup, the fusion center designs queries for each of the $M$ players in sequence and refines the posterior belief of the target location given the response of each player (see Figure \ref{fig:seq_scheme}). Recall the sub-time scale of sub-instants $\{n_t:t=0,\dots,M-1\}$ for each time instant $n$ and consider the filtration $\mathcal{G}_{n,t}$ defined in Section \ref{sec: multiplayer_entropy_loss}.A. Assuming that all players are queried in sequence starting from $m=1$ and ending at $m=M$, the posterior updates (after querying the $(t+1)$th player) become:
\begin{align*}
	p_{n_{t+1}}(x,\bepsilon) &= \PP(Y_{n_{t+1}}=y_{n_{t+1}}|A_{n_t},X^*=x,\epsilon_{t+1}^*=\epsilon_{t+1},\mathcal{G}_{n,t}) \\
	&\times p_{n_t}(x,\bepsilon) \\
	\PP&(Y_{n_{t+1}}=y_{n_{t+1}}|A_{n_t},X^*=x,\epsilon_{t+1}^*=\epsilon_{t+1},\mathcal{G}_{n,t}) \\
	&= \Bigg\{\begin{array}{ll} f_1^{(t+1)}(y_{n_{t+1}}|\epsilon_{t+1}),& x\in A_{n_t} \\ f_0^{(t+1)}(y_{n_{t+1}}|\epsilon_{t+1}),& x\notin A_{n_t} \end{array}. 
\end{align*}

\subsection{Joint Query Design}
In the joint setup, we assume that the fusion center designs queries for the $M$ players at each time instant $n$ and after querying all players, the responses are fused by the controller and the next set of questions is formulated. Recall the  filtration $\mathcal{F}_n$ defined in Section \ref{sec: multiplayer_entropy_loss}.B.

Define the density parametrized by $\bepsilon=(\epsilon_1,\dots,\epsilon_M)$ and $i_1,\dots,i_M \in \{0,1\}$:
\begin{equation*}
	g_{i_1:i_M}(\by|\bepsilon) = \prod_{m=1}^M f_{i_m}^{(m)}(y^{(m)}|\epsilon_m).
\end{equation*}

At the $n$th time instant, the posterior update becomes:
\begin{align*}
	p_{n+1}(x,\bepsilon) &= \PP(\bY_{n+1}=\by_{n+1}|\bA_n,X^*=x,\bepsilon^*=\bepsilon,\mathcal{F}_n) \\
	&\times p_n(x,\bepsilon) \\
	\PP&(\bY_{n+1}=\by_{n+1}|X^*=x,\bepsilon^*=\bepsilon,\mathcal{F}_n) \\
	&= \prod_{m=1}^M \Bigg\{ \begin{array}{ll} f_1^{(m)}(y_{n+1}^{(m)}|\epsilon_m), & x \in A_n^{(m)} \\ f_0^{(m)}(y_{n+1}^{(m)}|\epsilon_m), & x\notin A_n^{(m)} \end{array}.
\end{align*}

\subsection{Equivalence Theorems}
Since the error probabilities of the players are unknown, the joint policy derived in Theorem \ref{thm:joint} is no longer applicable or valid. The next theorem derives the jointly optimal policy for all players under the unknown channel case.

\begin{theorem} (Jointly Optimal Policy, Unknown Error Probabilities) \label{thm:unknown_joint}
	Let Assumptions \ref{assump:cond_indep_eps}, \ref{assump:BSC} and \ref{assump:query_region_regularity} hold. Consider the problem (\ref{eq: min_entr_obj}), where the joint policy is made up of the query regions for the $M$ players.
	\begin{enumerate}[leftmargin=*]
	\item Optimal policies $\bA_n=(A_n^{(1)},\dots,A_n^{(M)})$ at time $n$ satisfy:
	\begin{align}
		&G_n^* \nonumber \\
		&= \sup_{\bA} \Big\{ H\left( \sum_{i_1:i_M=0}^1 \int_{\bepsilon=0}^{1/2} g_{i_1:i_M}(\cdot|\bepsilon) P_n\left(\bigcap_{m} (A^{(m)})^{i_m},\bepsilon\right) d\bepsilon \right) \nonumber \\
			&\quad - \sum_{i_1:i_M=0}^1 \int_{\bepsilon=0}^{1/2} H(g_{i_1:i_M}(\cdot|\bepsilon)) P_n\left(\bigcap_{m} (A^{(m)})^{i_m},\bepsilon\right) d\bepsilon \Big\}. \label{eq:gain_joint}
	\end{align}
	\item The maximum information gain at time $n$ is:
		\begin{equation} \label{eq:G_optimal}
			G_n^* = \sum_{m=1}^M \EE[C(\epsilon_m)|\mathcal{F}_n],
		\end{equation}
		where $\EE[C(\epsilon_m)|\mathcal{F}_n]=\int_{\epsilon_m=0}^{1/2} C(\epsilon_m) p_n(\epsilon_m) d\epsilon_m$.
	\end{enumerate}
\end{theorem}
Next, we show a version of the equivalence theorem (Theorem \ref{thm:separation}) for the unknown channel case. This result is interesting since it implies that on average the joint query design is equivalent to the sequential query design, even when the error probabilities are unknown.
\begin{theorem} (Equivalence, Unknown Error Probabilities) \label{thm:equivalence_unknown}
	Let Assumptions \ref{assump:cond_indep_eps} and \ref{assump:BSC} hold. Consider the sequential and joint schemes described in Section \ref{sec: unknown_error_probability}.B and Section \ref{sec: unknown_error_probability}.C. \footnote{For the one-dimensional case, the sequential scheme implements (\ref{eq:gain_1d}) for each sub-instant to design a question for each player and the posterior is updated in sequence (see Figure \ref{fig:seq_scheme}).} Then, it follows that $G_{seq,n}^*=\EE[\sum_m C(\epsilon_m)|\mathcal{G}_n]$ and $G_n^* = \EE[\sum_m C(\epsilon_m)|\mathcal{F}_n]$ for all $n$.
\end{theorem}

\subsubsection{Lower Bound on MSE Performance}
The maximum entropy loss derived in Theorem \ref{thm:unknown_joint} is used next to provide a lower bound on the MSE of the joint sequential estimator.
\begin{theorem} (Lower bound on Joint MSE) \label{eq:lb_unknown}
Assume $H(p_0)$ is finite. Then, the joint MSE of the joint query policy in Theorem \ref{thm:unknown_joint} satisfies:
\begin{equation}
	\frac{K}{2\pi e} d \exp\left(-\frac{2n \bar{C}_n}{d}\right) \leq \EE[\nn X_n-X^*\nn_2^2] + \EE[\nn\bepsilon_n-\bepsilon^*\nn_2^2],
\end{equation}
where $K = \exp(2H(p_0))$ is a constant and $X_n=\EE[X^*|\mathcal{F}_n], \bepsilon_n = \EE[\bepsilon^*|\mathcal{F}_n]$. The expected entropy loss per iteration is $\bar{C}_n = \frac{1}{n}\sum_{k=0}^{n-1} G_k^*$.
\end{theorem}
\begin{IEEEproof}
	The proof follows using the result of part 2 of Theorem \ref{thm:unknown_joint} and similar bounding arguments as Theorem \ref{thm:lb}.
\end{IEEEproof}

\subsection{Discussion}
The jointly optimal policy derived for the unknown probability case in Theorem \ref{thm:unknown_joint} is reminiscent of the jointly optimal policy of Theorem \ref{thm:joint}. We remark that in the unknown probability setting, the maximum entropy loss $G_n^*$ given in (\ref{eq:gain_joint}) is not time-invariant, unlike in the case of known probability, in which the maximum entropy loss was the sum of the capacities of the players' channels (\ref{eq:joint_optimal}) and (\ref{eq:entropy_loss}). This observation motivates a player selection scheme; if we have the hard constraint that only one player may be used at a time, then, unlike in the known probability case, it may be that at different times, the maximal information gain may be obtained by different players.

\subsection{Player Selection Scheme}
We assume that at each time instant, only one player can be queried. We assume that the control $u_n=u$ implies that the $u$th player is to be queried at time $n$ and $A_n^{(u)}=A$ is the associated query region. Similarly to (\ref{eq:Bayes}), the joint posterior update in this case becomes:
\begin{align*}
	p_{n+1}(x,\bepsilon) &\propto \PP(Y_{n+1}^{(u)}|u_n=u,A_n^{(u)},X^*=x,\epsilon_u^*=\epsilon_u) \\
	&\qquad \times p_n(x,\bepsilon) \\
	\PP(Y_{n+1}^{(u)}=y^{(u)} &|u_n=u,A_n^{(u)},X^*=x,\epsilon_u^*=\epsilon_u) \\
	&= \left\{ \begin{array}{ll} f_1^{(u)}(y^{(u)}|\epsilon_u), & x\in A_n^{(u)} \\ f_0^{(u)}(y^{(u)}|\epsilon_u), & x\notin A_n^{(u)} \end{array} \right .
\end{align*}

\begin{theorem} (Player Selection Policy, Unknown Error Probabilities) \label{thm:unknown}
	Consider the problem (\ref{eq: min_entr_obj}), where the policy consists of which player to choose and the associated query region. At each time $n$:
	\begin{enumerate}[leftmargin=*]
	\item All optimal query policies satisfy:
	\begin{align}
		& \max_{u\in \{1,\dots,M\}} G_n^*(u) = \sup_{A} \Big\{ H\Bigg( \int_{\epsilon_u=0}^{1/2} f_1(\cdot|\epsilon_u) P_n^{(u)}(A,\epsilon_u) \nonumber \\
		&\qquad + f_0(\cdot|\epsilon_u) P_n^{(u)}(A^c,\epsilon_u) d\epsilon_u \Bigg) \nonumber \\
		&- \int_{\epsilon_u=0}^{1/2} \Big( H\left(f_1(\cdot|\epsilon_u)\right) P_n^{(u)}(A,\epsilon_u) \nonumber \\
		&\qquad + H\left(f_0(\cdot|\epsilon_u)\right) P_n^{(u)}(A^c,\epsilon_u) \Big) d\epsilon_u  \Big\}.		 \label{eq:gain}
	\end{align}
	\item The maximum entropy loss is:
		\begin{equation*}
			G_n^* = \max_{u} G_n^*(u) = \max_u \EE[C(\epsilon_u)|\mathcal{F}_n].
		\end{equation*}
	\end{enumerate}
\end{theorem}
The optimal policy for the minimum expected entropy criterion (\ref{eq: min_entr_obj}) shown in Theorem \ref{thm:unknown} is intuitive. The player $u$ with the maximum information gain (or entropy loss) is chosen, where the entropy loss is measured as a function of the $u$-th sub-marginal distribution $p_n^{(u)}(x,\epsilon_u)$. While the form (\ref{eq:gain}) bears some similarity to the form (\ref{eq:joint_optimal}), the bisection policy is no longer optimal. In addition, in this unknown probability setting, it may not always be the case that the player with the largest capacity will be chosen (this is the case in the known probability setting). The integral over $\epsilon\in [0,1/2)$ essentially averages out the contribution of the unknown error probabilities with respect to the observed data up to the current time $n$. 

\subsubsection{One-dimensional Case}
The next corollary specifies the form of the optimal policy derived in Theorem \ref{thm:unknown} for one-dimensional targets. For simplicity, consider the unit interval $\mathcal{X}=[0,1]$ as the target domain.

\begin{corollary} (Player Selection Policy, Unknown Error Probabilities, One-dimensional Target) \label{cor:unknown}
	Consider the problem (\ref{eq: min_entr_obj}) for the optimal player and query selection policy. Consider the query regions $A_n=[0,x_n]$. The optimal player $u$ and associated query region $A=[0,x]$ at time $n$ is given by:
	\begin{equation} \label{eq:gain_1d}
		\max_{u} \left\{  \max_{x \in [0,1]} h_B(g_{1,n}^{(u)}(x)) - c_n^{(u)} \right\},
	\end{equation}
	where $h_B(\cdot)$ is the binary entropy function \cite{CoverThomas} and
	\begin{align*}
		c_n^{(u)} &= \int_{\epsilon_u=0}^{1/2} h_B(\epsilon_u) p_n^{(u)}(\epsilon_u) d\epsilon_u \\
		g_{1,n}^{(u)}(x) &= \int_0^x \mu_n^{(u)}(t) dt + \int_x^1 (p_n(t)-\mu_n^{(u)}(t)) dt \\
		\mu_n^{(u)}(t) &= \int_{\epsilon_u=0}^{1/2} \epsilon_u p_n^{(u)}(t,\epsilon_u)  d\epsilon_u \\
		p_n(t) &= \int_{\epsilon_1=0}^{1/2} \cdots \int_{\epsilon_M=0}^{1/2} p_n(t,\epsilon_1,\dots,\epsilon_M) d\epsilon_1 \cdots d\epsilon_M .
	\end{align*}
\end{corollary}
We note that the optimal policy derived for the unknown probability case in (\ref{eq:gain_1d}) is \textit{not} equivalent to the probabilistic bisection policy-i.e., obtaining $P_n^{(u)}([0,x_n^{(u)}])=1/2$ for each player $u$ and then evaluating the information gain and choosing the player with the maximum information gain. This heuristic scheme would yield a suboptimal information gain as compared to the maximal information gain given by (\ref{eq:gain_1d}). Thus, in the unknown probability setting, the optimal control law is no longer equivalent to the known probability setting (after marginalizing out the noise parameters $\epsilon_1,\dots,\epsilon_M$). This result shows that the two settings are quite different and the answers to the unknown channel case are more complex. We empirically observed that there is a unique query point $x=x_n^*=x_n^{(u^*)}$ that maximizes the function (\ref{eq:gain_1d}). This is similar to the one-dimensional case for the known probability setting when the query region is of the form $A=[0,x]$; i.e., the optimal point is the median.

\section{Simulations} \label{sec: simulations}
This section contains a few illustrative simulations that validate the methodology presented throughout the paper. In all simulations, MATLAB R2012b was used.

\subsection{Known Error Probability}
\begin{figure*}[ht]
	\centering
		\includegraphics[width=0.80\textwidth]{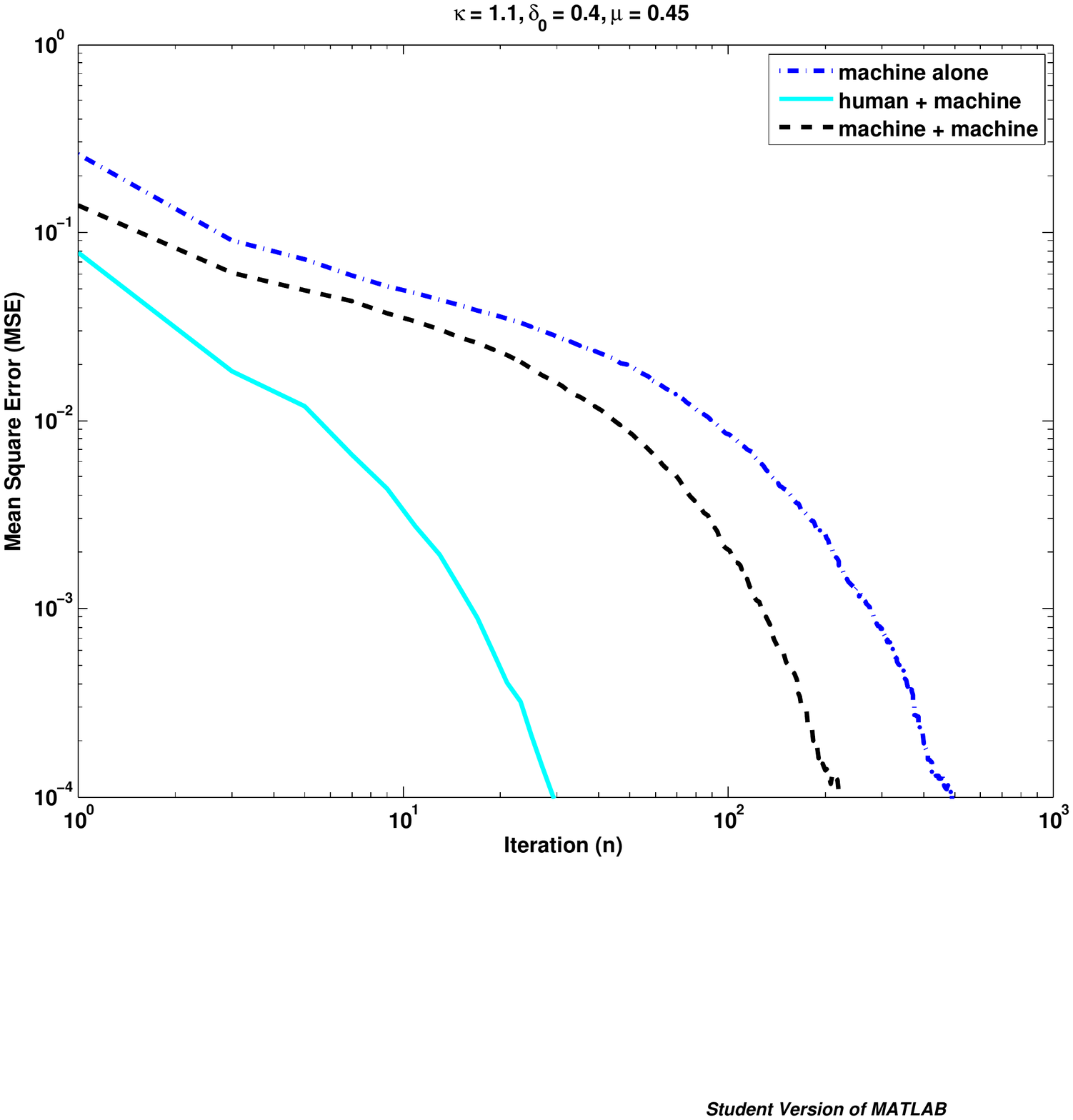}
	\caption{Monte Carlo simulation for MSE performance of the sequential estimator as a function of iteration for a single player 20 questions game (machine alone), a two player game without human-machine collaboration (machine+machine), and a two player game with human-machine collaboration (human+machine). 8000 Monte Carlo trials were used. The human player's parameters were set to $\kappa=1.1, \mu=0.45, \delta_0=0.4$, the machine players' parameters were $\epsilon_1=\epsilon_2=0.4$, and the length of the pseudo-posterior was $\Delta^{-1}=1500$. The target location was set to $X^*=0.75$. The initial distribution was uniform.}
	\label{fig:emp_MSE_uniform}
\end{figure*}

\begin{figure*}[ht]
	\centering
		\includegraphics[width=0.80\textwidth]{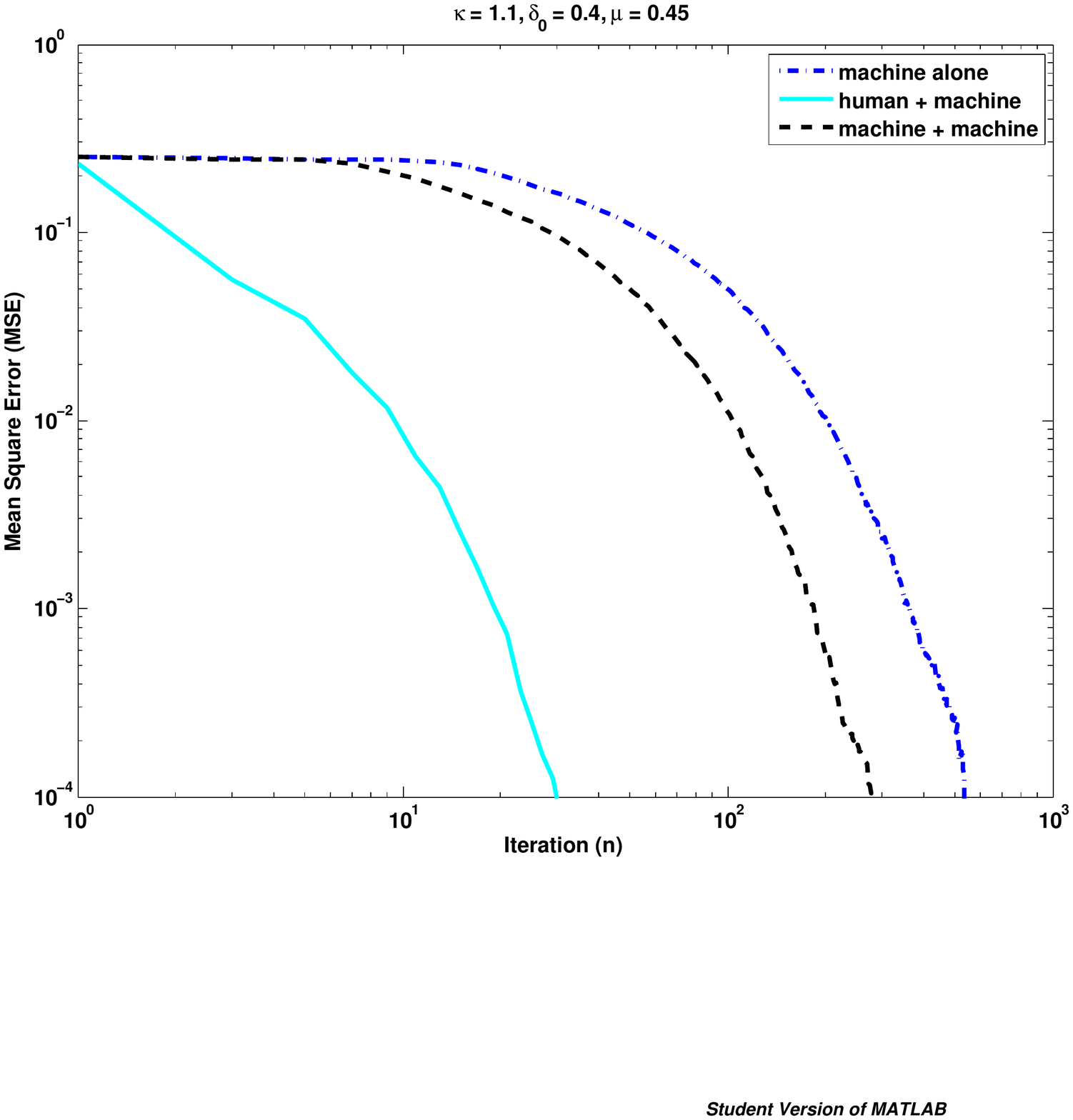}
	\caption{Monte Carlo simulation for MSE performance of the sequential estimator as a function of iteration for a single player 20 questions game (machine alone), a two player game without human-machine collaboration (machine+machine), and a two player game with human-machine collaboration (human+machine). 8000 Monte Carlo trials were used. The human player's parameters were set to $\kappa=1.1, \mu=0.45, \delta_0=0.4$, the machine players' parameters were $\epsilon_1=\epsilon_2=0.4$, and the length of pseudo-posterior was $\Delta^{-1}=1500$. The target was set to $X^*=0.75$. The initial distribution was a mixture of three Gaussian distributions as shown in Figure \ref{fig:mixture}.}
	\label{fig:emp_MSE_nonuniform}
\end{figure*}

Figures \ref{fig:emp_MSE_uniform} and \ref{fig:emp_MSE_nonuniform} show the empirical performance of the human-in-the-loop by comparing the mean-square error (MSE) of target localization for three scenarios: a 20 questions game with only a single player (machine alone); a 20 questions game with two machine players (machine + machine); and a 20 questions game with one machine and one human player (human + machine).

Figure \ref{fig:emp_MSE_uniform} and \ref{fig:emp_MSE_nonuniform} show the MSE for the respective cases of uniform and nonuniform prior distributions on the target location. The BSC crossover probability for the machine was set to $\epsilon_1=0.4$ while the BSC crossover probability for the human were determined by (\ref{eq:human_err}) with $\kappa=1.1, \delta_0=0.4$ and $\mu=0.45$. These are the same parameter settings as used to generate the blue curve shown in Figure \ref{fig:human_error_model}, where the human is significantly more accurate than the machine in initial iterations while the opposite is true in the final iterations. A total of $8000$ Monte Carlo runs were averaged to generate the curves in Figures \ref{fig:emp_MSE_uniform} and \ref{fig:emp_MSE_nonuniform}. Each Monte Carlo run consists of implementing the sequential query design via the BZ algorithm with a discretization of the interval $[0,1]$ into $1500$ equal cells representing possible target positions. The target position was set to $X^*=0.75$, the BSC error channel was simulated using biased coin flipping and the posterior distributions were updated at every iteration according to (\ref{eq:Bayes}). We note that BZ algorithm was used to implement each bisection, and no query costs or sensor selection were used. In this setting, we assumed that the BSC crossover probabilities are known to the controller.

It is observed that employing a human in the loop reduces the MSE for a wide range of $n$. We note that as $n\to\infty$, the machine + human curve will cross the machine + machine curve, being consistent with the upper bounds shown in (\ref{eq:ub}) and (\ref{eq:ub_human}) since the human's contribution is strongest in the first few iterations, while its value decreases to zero as $n\to\infty$. Also, note that the human model does not yield a different exponent in the exponential rate of convergence, while adding a second player does (as predicted in Theorems \ref{thm:lb} and \ref{thm:ub}).

\begin{figure}[ht]
	\centering
		\includegraphics[width=0.40\textwidth]{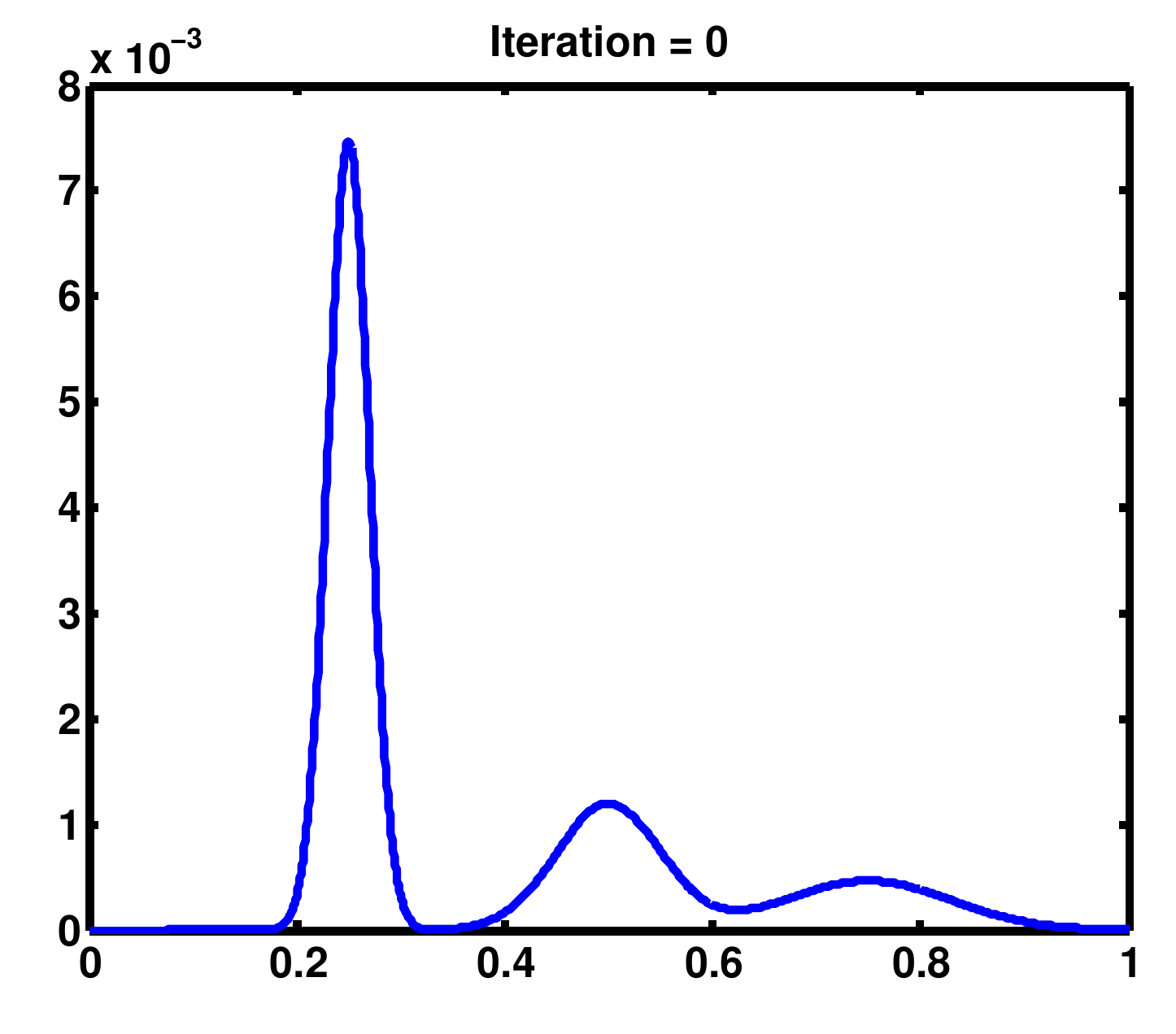}
	\caption{Initial distribution for BZ algorithm. The distribution is a mixture of three Gaussians with means $0.25$, $0.5$ and $0.75$, and variances $0.02$, $0.05$ and $0.08$, respectively. The target was set to be the center of the mode at $X^*=0.75$ with the largest variance. The resulting MSE performance of the sequential estimator is shown in Figure \ref{fig:emp_MSE_nonuniform}.}
	\label{fig:mixture}
\end{figure}

Next, we observe the effect of the prior distribution associated with the target location on the MSE performance. We observe that the machine + human provides a larger gain when the initial distribution is trimodal with larger variance on the true component centered at $X^*=0.75$ (see Figure \ref{fig:mixture}) as shown in Figure \ref{fig:emp_MSE_nonuniform}, as compared to the gain when starting from a uniform distribution as shown in Figure \ref{fig:emp_MSE_uniform}. In fact, the human-in-the-loop combined with a machine outperforms two machines for a wide range of iterations $n$.

\begin{figure*}[ht]
	\centering
		\includegraphics[width=0.80\textwidth]{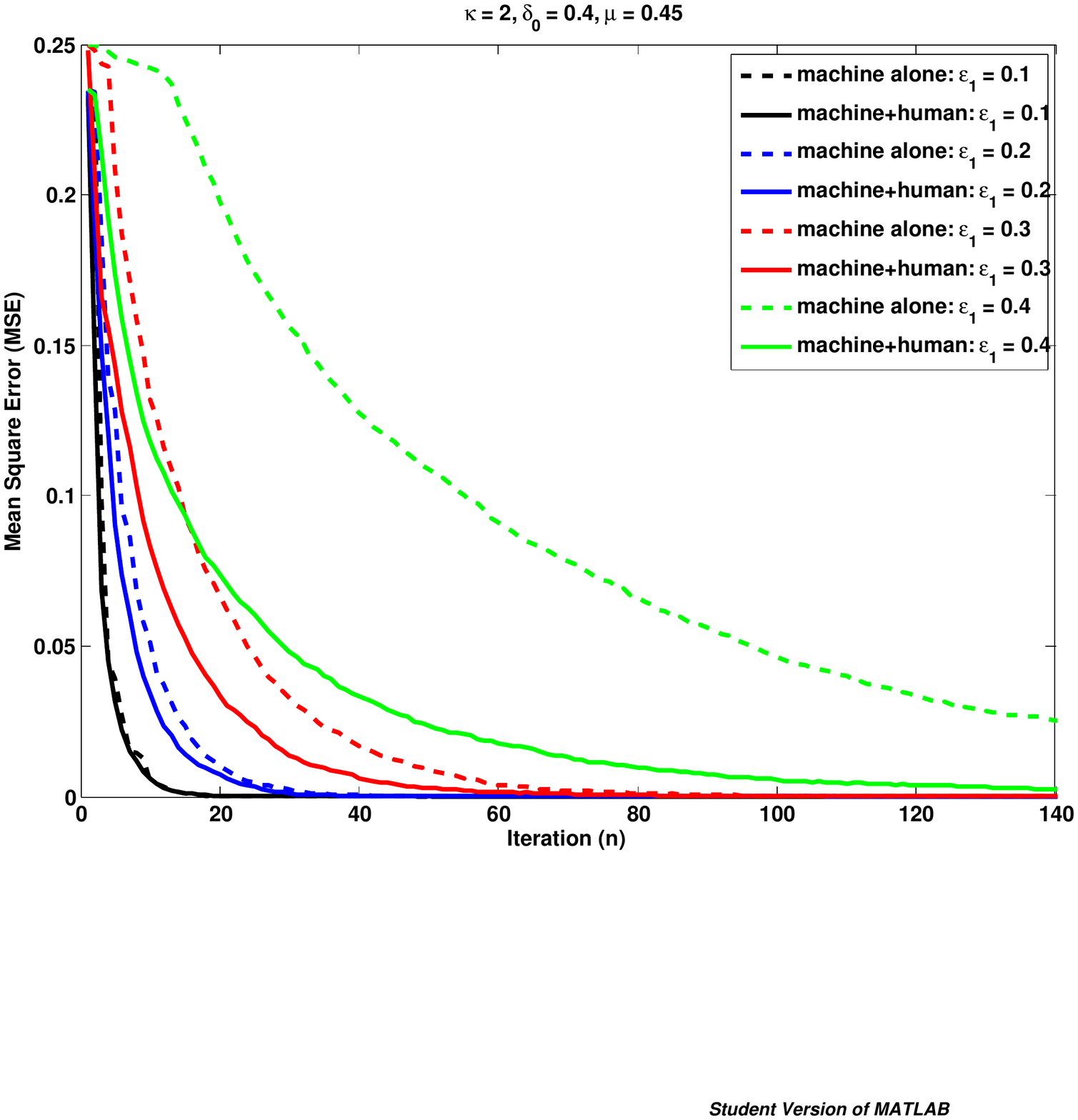}
	\caption{Monte Carlo simulation for MSE performance of the sequential estimator as a function of iteration and $\epsilon_1\in(0,1/2)$. 2000 Monte Carlo trials were used. The human parameters were set to $\kappa=2.0, \mu=0.45, \delta_0=0.4$, the length of pseudo-posterior was $\Delta^{-1}=1500$. The target was set to $X^*=0.75$. The initial distribution was a mixture of three Gaussian distributions as shown in Figure \ref{fig:mixture}.}
	\label{fig:emp_MSE_kap20}
\end{figure*}

\begin{figure*}[ht]
	\centering
		\includegraphics[width=0.80\textwidth]{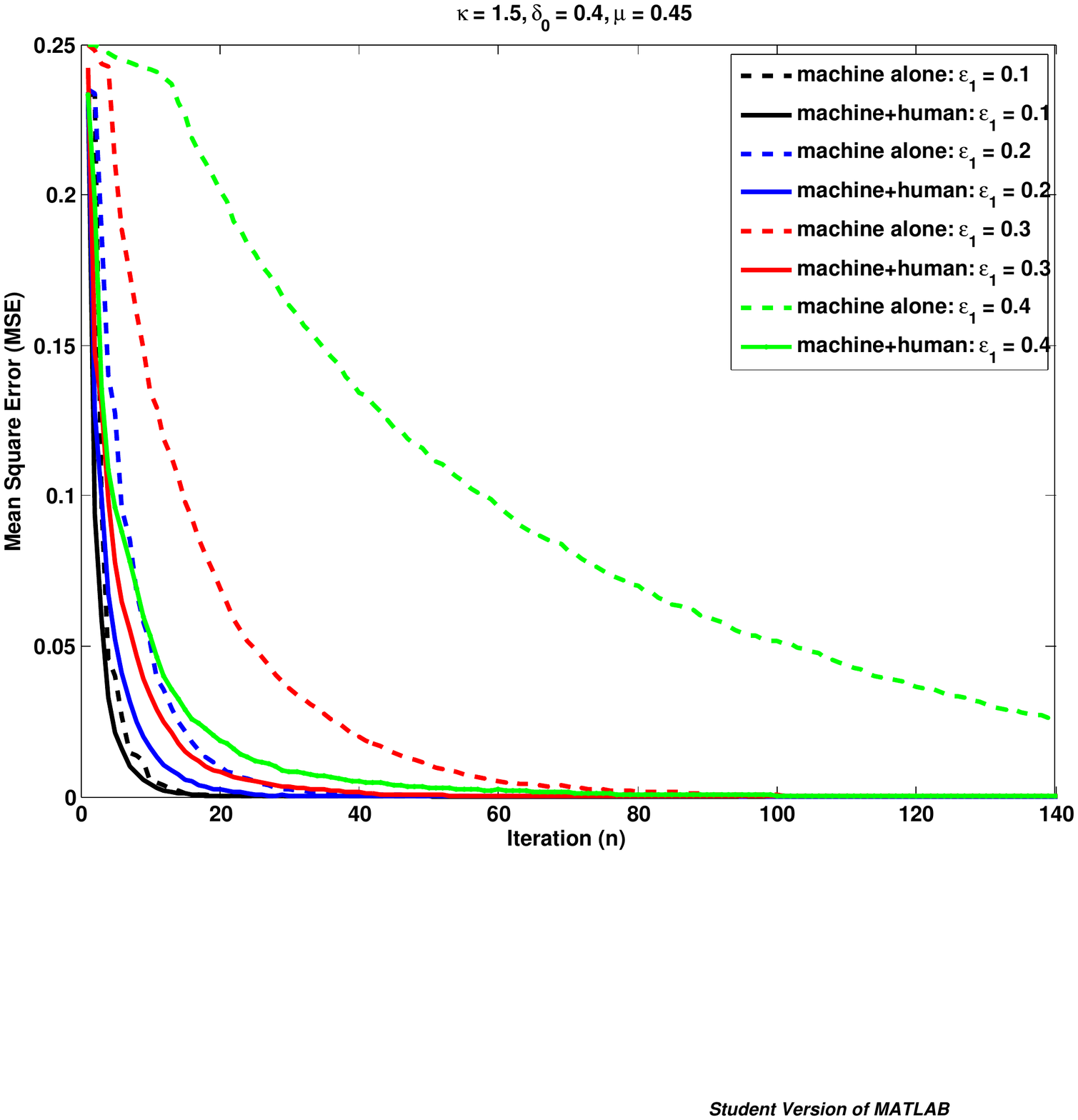}
	\caption{Monte Carlo simulation for MSE performance of the sequential estimator as a function of iteration and $\epsilon_1\in(0,1/2)$. 2000 Monte Carlo trials were used. The human parameters were set to $\kappa=1.5, \mu=0.45, \delta_0=0.4$, the length of pseudo-posterior was $\Delta^{-1}=1500$. The target was set to $X^*=0.75$. The initial distribution was a mixture of three Gaussian distributions as shown in Figure \ref{fig:mixture}.}
	\label{fig:emp_MSE_kap15}
\end{figure*}

Figures \ref{fig:emp_MSE_kap20} and \ref{fig:emp_MSE_kap15} show the empirical MSE as a function of $\epsilon_1\in (0,1/2)$ for $\kappa=2.0$ and $\kappa=1.5$, respectively. As expected, larger MSE gains are obtained for $\kappa=1.5$. For fixed $\kappa$, we observe from both figures that the MSE associated with the machine alone increases as $\epsilon_1$ increases, and in addition, the MSE associated with ``machine + human'' yields a larger improvement over just using the machine for larger $\epsilon_1$. In other words, the worse the machine is, the larger the value of the human in reducing the MSE.

\subsection{Unknown Error Probability}
\begin{figure}[ht]
	\centering
		\includegraphics[width=0.50\textwidth]{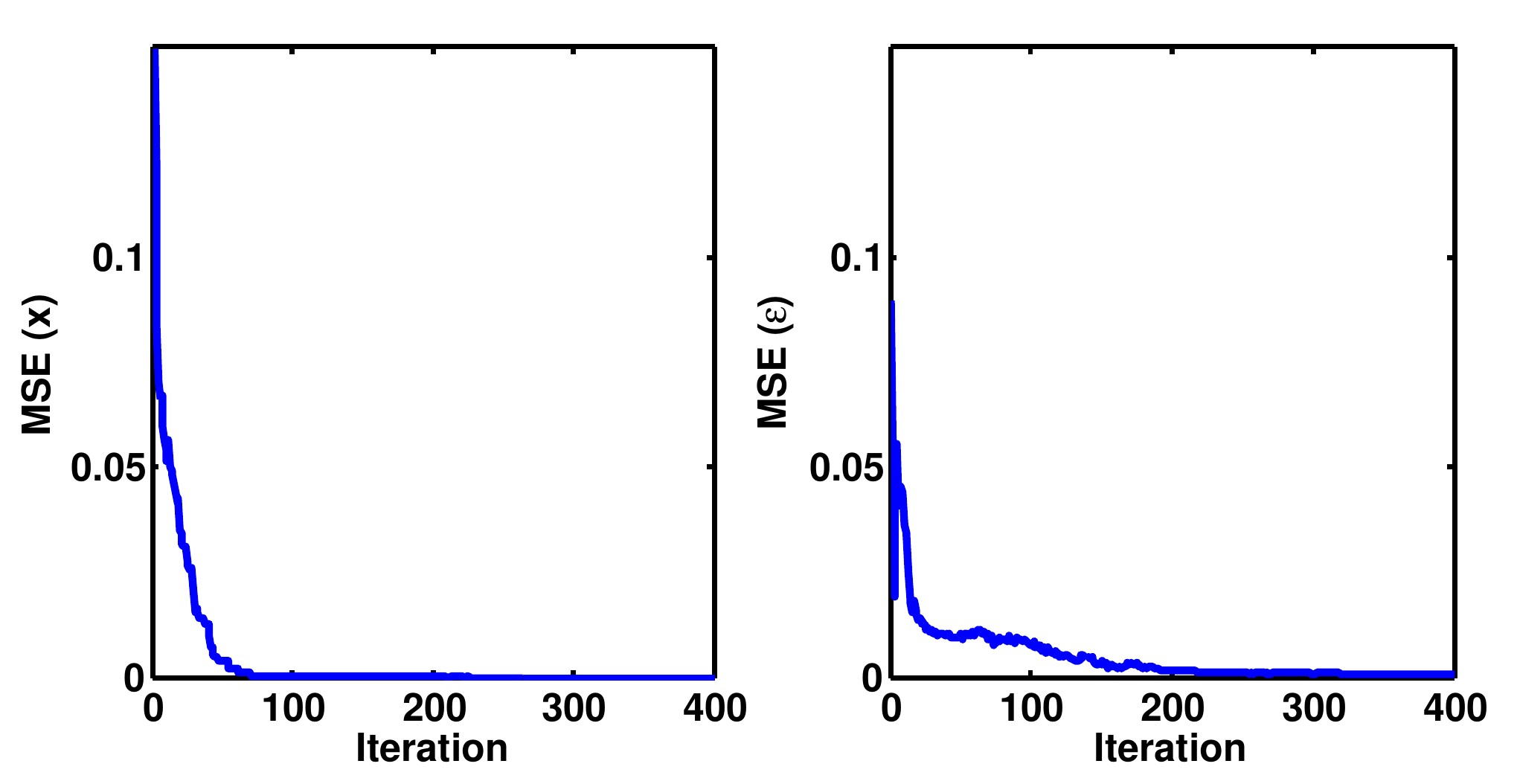}
	\caption{ Monte Carlo simulation for MSE performance of the joint sequential estimator (of the target $X^*$ and the error probability $\epsilon^*$). The MSE for target location $X$ is shown on the left and MSE for BSC crossover probability $\epsilon$ on the right, as a function of iteration. Note that both decay to zero over iteration. Interestingly, the target location estimator achieves nearly zero MSE well before that of the channel estimation error, indicating a certain robustness of estimates of $X^*$ to errors in estimates of $\epsilon$. 100 Monte Carlo trials were used. The true error probability was set to $\epsilon^*=0.3$ and the true target location was $X^*=0.75$. The initial distribution was a joint uniform density $p_0(x,\epsilon)$. }
	\label{fig:emp_MSE_unknown}
\end{figure}

Figure \ref{fig:emp_MSE_unknown} numerically evaluates the MSE performance for $M=1$ player with unknown error probability. This simulation implies that the binary responses obtained from one player carry enough information to accurately estimate the target and its error probability.

\section{Conclusion} \label{sec: conclusions}
We studied the problem of collaborative 20 questions with noise for the multiplayer case. We derived an equivalence theorem that shows the joint query design has the same performance on average as the sequential bisection query design, despite the fact that the sequential bisection query design has access to a more refined filtration. In addition, the sequential bisection query design is easily implemented due to the low complexity of the controllers (unlike the jointly optimal design). Using this framework, we obtained mean-square-error bounds for the performance of the sequential estimator. The methodology was applied to human-in-the-loop target localization systems. 

The framework was generalized to the case of unknown error probabilities associated with noisy players. For this case, it was shown that the maximum entropy loss per iteration is time-varying (unlike in the known probability case) and the optimal policy that achieves this gain is not equivalent to the probabilistic bisection policy. Simulations were provided to numerically evaluate the performance of the proposed sequential estimator. Worthwhile future work could include the following extensions: 1) query design for target detection and classification; 2) more sophisticated query-response models that account for state-dependent response (channel) errors; 3) optimal query design that restricts the complexity of the questions, e.g., to half planes as in \cite{NowakGBS:2011}; 4) extension of theory to the case where there are additional query-dependent costs on acquisition of information from each player.

\section*{Acknowledgment}
The authors gratefully acknowledge the insightful comments of several reviewers that contributed to improved readability of this paper. 

\appendices

\section{Proof of Theorem \ref{thm:joint}}

\begin{IEEEproof}
  Using (\ref{eq:indep}) and (\ref{eq:exp}), we have:
	\begin{align}
		&\PP(\bY_{n+1}=\by|\bA_n,X^*=x,\mathcal{F}_n) \nonumber \\
			&= \prod_{m=1}^M \Big\{ f_1^{(m)}(y^{(m)}|A_n^{(m)},\mathcal{F}_n) I(x \in A_n^{(m)}) \nonumber \\
			&\quad + f_0^{(m)}(y^{(m)}|A_n^{(m)},\mathcal{F}_n) I(x\notin A_n^{(m)}) \Big\}  \nonumber \\
			&= \sum_{i_1:i_M=0}^1 g_{i_1:i_M}(\by|\bA_n,\mathcal{F}_n) I\left(x\in \bigcap_{m=1}^M (A_n^{(m)})^{i_m}\right). \label{eq:prob1}
	\end{align}
	By integrating over $x\in \mathcal{X}$, we have:
	\begin{align}
		\PP&(\bY_{n+1}=\by|\bA_n,\mathcal{F}_n) = \EE[\PP(\bY_{n+1}=\by|\bA_n,X^*,\mathcal{F}_n)] \nonumber \\
			&= \sum_{i_1:i_M=0}^1 g_{i_1:i_M}(\by|\bA_n,\mathcal{F}_n) P_n\left(\bigcap_{m=1}^M (A_n^{(m)})^{i_m}\right). \label{eq:prob2}
	\end{align}
	The expected one step entropy loss is related to the conditional mutual information as:
	\begin{align}
		H(p_n) &- \EE[H(p_{n+1})|\bA_n,\mathcal{F}_n] = I(X^*; \bY_{n+1}|\bA_n,\mathcal{F}_n) \nonumber \\
			&= H(\bY_{n+1}|\bA_n,\mathcal{F}_n) - \EE[H(\bY_{n+1})|X^*,\bA_n,\mathcal{F}_n]. \label{eq:mutual_info}
	\end{align}
	From (\ref{eq:prob2}), we have:
	\begin{equation*}
		H(\bY_{n+1}|\bA_n,\mathcal{F}_n) = H\left( \sum_{i_1:i_M=0}^1 g_{i_1:i_M}(\cdot) P_n\left(\bigcap_{m=1}^M (A_n^{(m)})^{i_m}\right) \right)
	\end{equation*}
	and using (\ref{eq:prob1}):
	\begin{align*}
		\EE &[H(\bY_{n+1})|X^*,\bA_n,\mathcal{F}_n] \\
			&= \int_{\mathcal{X}} p_n(x) H(\bY_{n+1}|X^*=x,\bA_n,\mathcal{F}_n) dx \\
			&= \sum_{i_1:i_M=0}^1 H\left(g_{i_1:i_M}\right) P_n\left(\bigcap_{m=1}^M (A_n^{(m)})^{i_m}\right).
	\end{align*}
	Thus, taking the supremum of both sides in (\ref{eq:mutual_info}), we conclude that the entropy can be reduced at most by $G^*$, i.e.,
	\begin{equation} \label{eq:max_entropy_loss}
		\sup_{\bA} \left\{H(p_n) - \EE[H(p_{n+1})|\bA_n=\bA,\mathcal{F}_n] \right\} = G^*
	\end{equation}
	where $G^*$ is defined in (\ref{eq:joint_optimal}).
		
	Consider the optimal control problem (\ref{eq: min_entr_obj}), and define the value function:
	\begin{equation} \label{eq:value_function}
		V_n(p) = \inf_\zeta \EE^\zeta[H(p_N)|p_n=p], \quad n=0,\dots,N
	\end{equation}
	It is well known from stochastic control theory that the value function (\ref{eq:value_function}) satisfies Bellman's recursion (\cite{Dynkin:1979}):
	\begin{equation} \label{eq:Bellman_recursion}
		V_n(p) = \inf_{\bA} \EE[V_{n+1}(p_{n+1})|\bA_n=\bA,p_n=p], \quad n<N
	\end{equation}
	and any policy attaining the infimum in (\ref{eq:Bellman_recursion}) is optimal. To finish the proof, we find an explicit form for the value function and show that the policy achieving the supremum (\ref{eq:joint_optimal}) achieves the infimum in (\ref{eq:Bellman_recursion}). We remark that the function $V_n(\cdot)$ is Borel measurable since the entropy functional $H(\cdot)$ is measurable under on the query regions $\bA_n$ \cite{Jedynak12}. We claim that the value function (\ref{eq:value_function}) is given by:
	\begin{equation} \label{eq:value_function_explicit}
		V_n(p_n) = H(p_n) - (N-n)G^*, \quad n=0,\dots,N
	\end{equation}
	We proceed by backward induction. The base case is trivial since $V_N(p_N)=H(p_N)$. Fix an arbitrary $n<N$ and assume that $V_{n+1}(p_{n+1})=H(p_{n+1})-(N-(n+1))G^*$. Then, from (\ref{eq:Bellman_recursion}) and the induction hypothesis:
	\begin{align*}
		V_n(p_n) &= \inf_\bA \EE[V_{n+1}(p_{n+1})|\bA_n=\bA,p_n] \\
			&= \inf_\bA \EE[H(p_{n+1})|\bA_n=\bA,p_n] - (N-n-1)G^* \\
			&= H(p_n) - G^* - (N-n)G^* + G^* \\
			&= H(p_n) - (N-n) G^*
	\end{align*}
	where we used (\ref{eq:max_entropy_loss}). It follows that the optimal query set $\bA_n$ must satisfy (\ref{eq:joint_optimal}).	
\end{IEEEproof}

\section{Proof of Theorem \ref{thm:separation}}

\begin{IEEEproof}
	Let $G_{seq}$ denote the maximum expected entropy loss after querying $M$ players sequentially. The bisection policy yields an expected entropy loss of $C(\epsilon_m)=1-h_b(\epsilon_m)$ \footnote{This is the capacity of the $m$th BSC \cite{Jedynak12,CoverThomas}.} after querying the $m$th player \cite{Jedynak12}. Thus, $G_{seq}=\sum_{m=1}^M C(\epsilon_m)$. The expected entropy loss at sub-time instant $n_t$ is $H(p_{n_t})-\EE[H(p_{n_{t+1}})|A_{n_t},\mathcal{G}_{n_t}] = I(X^*;Y_{n_{t+1}}|A_{n_t},\mathcal{G}_{n_t})$. To show this rigorously, observe:
	\begin{align*}
		&G_{seq} \\
			&= \sup_{\{A_{n_t}\}_{t=0}^{M-1}} \EE[H(p_{n})-H(p_{n+1})|\mathcal{G}_n] \\
			&= \sup_{\{A_{n_t}\}_{t=0}^{M-1}} \EE\left[\sum_{t=0}^{M-1} H(p_{n_t})-H(p_{n_{t+1}}) \Bigg| \mathcal{G}_n  \right] \\
			&= \sup_{\{A_{n_t}\}_{t=0}^{M-1}} \sum_{t=0}^{M-1} \EE\left[ \EE\left[ H(p_{n_t})-H(p_{n_{t+1}}) \Bigg|A_{n_t}, \mathcal{G}_{n_t} \right] \Bigg| \mathcal{G}_n \right] \\
			&= \sup_{\{A_{n_t}\}_{t=0}^{M-1}} \EE\left[ \sum_{t=0}^{M-1} I(X^*;Y_{n_{t+1}}|A_{n_t},\mathcal{G}_{n_t}) \Bigg| \mathcal{G}_n \right] \\
			&= \EE\left[ \sum_{t=0}^{M-1} \sup_{A_{n_t}} I(X^*;Y_{n_{t+1}}|A_{n_t},\mathcal{G}_{n_t}) \Bigg| \mathcal{G}_n  \right] \\
			&= \EE\left[ \sum_{t=0}^{M-1} C(\epsilon_{t+1}) \Bigg| \mathcal{G}_n \right] = \sum_{m=1}^M C(\epsilon_m).
	\end{align*}
	To finish the proof, we show $G_{seq}=G^*$.
	The consequence $G_{seq}=G^*$ follows from the chain rule of conditional mutual information, but we show an argument based on convex optimization that characterizes the jointly optimal policy as well. From Theorem \ref{thm:joint},
	\begin{align}
		G^* &= \sup_{A^{(1)},\dots,A^{(M)}} \Big\{ H\left( \sum_{i_1:i_M=0}^1 g_{i_1:i_M}(\cdot) P_n\Big(\bigcap_{m=1}^M (A_n^{(m)})^{i_m}\Big) \right) \nonumber \\
		 &-\sum_{i_1:i_M=0}^1 H\left(g_{i_1:i_M}(\cdot)\right) P_n\Big(\bigcap_{m=1}^M (A_n^{(m)})^{i_m}\Big) \Big\} \nonumber \\
		 &= \sup_{\bp} \Big\{ H\left( \sum_{i_1:i_M=0}^1 g_{i_1:i_M}(\cdot) p_{i_1,\dots,i_M} \right) \nonumber \\
		 &- \sum_{i_1:i_M=0}^1 H(g_{i_1:i_M}(\cdot)) p_{i_1,\dots,i_M} : \bp\succeq 0, 1^T\bp=1 \Big\} \nonumber   \\
		 &= \sup_{\bp} \{H(\bp^T\bg) - \bp^TH(\bg): \bp\succeq 0, 1^T\bp = 1 \} \label{eq:max} \\
		 &= G_{seq} \nonumber
	\end{align}
	where the probability vector $\bp \in \RR^{2^M}$ contains $p_{i1,\dots,i_M}$, and $\bg$ contains the distributions $g_{i_1:i_M}(\cdot)$. The last equality follows by the symmetry of the BSC. The supremum in the strictly concave problem (\ref{eq:max}) is achieved by the uniform distribution. This is justified by noting that the second term is independent of $\bp$ since for $1^T\bp=1$, we obtain:
	\begin{align*}
		\bp^T H(\bg) &= \sum_{i_1:i_M=0}^1 H\left(\prod_{m=1}^M f_{i_m}^{(m)}(\cdot) \right) p_{i_1,\dots,i_M} \\
			&= \sum_{i_1:i_M=0}^1 \sum_{m=1}^M H\left(f_{i_m}^{(m)}(\cdot) \right) p_{i_1,\dots,i_M} \\
			&= \sum_{m=1}^M h_B(\epsilon_m) \cdot \sum_{i_1=0}^1 \cdots \sum_{i_M=0}^1 p_{i_1,\dots,i_M} \\
			&= \sum_{m=1}^M h_B(\epsilon_m).
	\end{align*}
	Thus, the supremum of (\ref{eq:max}) can be restricted to the first term which is achieved by $p_{i_1,\dots,i_M}^*=2^{-M}$ since:
	\begin{align*}
		&H\left(\sum_{i_1:i_M=0}^1 g_{i_1:i_M}(\by) p_{i_1,\dots,i_M}^* \right) \\
			&= H\left(2^{-M} \sum_{i_1=0}^1 \cdots \sum_{i_M=0}^1 \prod_{m=1}^M (1-\epsilon_m)^{I(y^{(m)}=i_m)} \epsilon_m^{1-I(y^{(m)}=i_m)} \right) \\ 
			&= H(u(\cdot)) = \log_2(2^M) = M
	\end{align*}
	where $u(\cdot)$ is the uniform distribution over $\{0,1\}^M$.
\end{IEEEproof}

\section{Proof of Theorem \ref{thm:policy_cost}}
\begin{IEEEproof}
	It suffices to show that for each $n=0,\dots,N$ that the value function is given by (\ref{eq:value_function_cost}) and that optimal policies satisfy Bellman's recursion. We do this by backward induction. We remark that the value function is measurable since the entropy functional is measurable \cite{Jedynak12}. The base case is trivial:
	\begin{equation*}
		V_N(p_N,T_N) = \EE[H(p_N)+\gamma T_N|p_N,T_N] = H(p_N) + \gamma T_N
	\end{equation*}
	Now, fixing $n<N$ and assuming that the value function is of the form (\ref{eq:value_function}) for $n+1$, we have by Bellman's recursion (\ref{eq:value_recursion}):
	\begin{align}
		V_n&(p_n,T_n) \nonumber \\
			&= \inf_{1\leq u\leq M,A} \EE[V_{n+1}(p_{n+1},T_{n+1}) | u_n=u, A_n=A, p_n, T_n] \nonumber \\
			&= \inf_{u, A} \EE\Big[H(p_{n+1})+\gamma T_{n+1} \nonumber \\
			&\quad - \sum_{k=n+1}^{N-1} \sup_{u',A'} G_k(u',A') \Big| u_n=u, A_n=A, p_n, T_n \Big] \nonumber \\
			\end{align}
			\begin{align}
			&= H(p_n) + \gamma T_n - \sum_{k=n+1}^{N-1} \sup_{u,A} G_k(u,A) \nonumber \\
			&\quad - \sup_{u,A} \left\{ I(X^*;\bY_{n+1}|u_n=u,A_n=A,p_n) - \gamma K_n(u,A) \right\} \nonumber \\
			&= H(p_n) + \gamma T_n - \sum_{k=n}^{N-1} \sup_{u,A} G_k(u,A) \label{eq:Bellman1}
	\end{align}
	where we used the identities:
	\begin{align*}
		\EE[H(p_{n+1})|u_n,A_n,p_n] &= H(p_n)-I(X^*;\bY_{n+1}|u_n,A_n,p_n) \\
		T_{n+1} &= T_n + K_n(u_n,A_n)
	\end{align*}
	The optimality condition of the policy follows from the condition that the infimum in Bellman's equation (\ref{eq:Bellman1}) is achieved.
\end{IEEEproof}

\section{Proof of Corollary \ref{cor:min_entropy_cost}}
\begin{IEEEproof}
	From Theorem \ref{thm:separation}, we have:
\begin{equation*}
	\inf_{A} \EE[H(p_{n+1}) | u_n=u,A_n=A,p_n] = H(p_n) - C_n(u)
\end{equation*}
Using this in the proof of Theorem \ref{thm:policy_cost}, it follows that the optimality condition (\ref{eq:optim_condition}) satisfies (\ref{eq:balance}).
\end{IEEEproof}

\section{Proof of Theorem \ref{thm:lb}}

\begin{IEEEproof}
	We note from the proof of Theorem \ref{thm:joint} or Theorem \ref{thm:separation}, for any policy $\zeta$, we have $\EE^\zeta[H(p_n)] \geq H(p_0) - nC$ \footnote{For optimal policies $\zeta$, this becomes an equality.}. Let $K_n$ denote the conditional error covariance of the random vector $e_n=X^*-\EE[X^*|\bY^n]$, i.e., $K_n=\Cov(e_n|\bY^n)$. From Theorem 17.2.3 in \cite{CoverThomas}	and Jensen's inequality, we have:
	\begin{align*}
		\EE^\zeta[H(p_n)] &\leq \EE^\zeta \left[ \frac{1}{2} \log((2\pi e)^d \det(K_n)) \right] \\
			&\leq \frac{1}{2}\log((2\pi e)^d) + \frac{1}{2} \log(\det(\EE^\zeta[K_n])) \\
			&= \frac{1}{2} \log((2\pi e)^d \det(\EE^\zeta[K_n]))
	\end{align*}
	where $\det(\cdot)$ denotes the determinant.	Rewriting this:
	\begin{align*}
		  \frac{K \exp(-2nC)}{(2\pi e)^d} &\leq \frac{\exp(2 \EE^\zeta[H(p_n)])}{(2\pi e)^d} \\
			&\leq \det(\EE^\zeta[K_n]) \leq \left(\frac{\EE^\zeta[\tr(K_n)]}{d}\right)^d
	\end{align*}
	where $\tr(\cdot)$ denotes the trace of a matrix. Note that we also used the inequality of arithmetic and geometric means in the last step. Using the fact that the conditional mean minimizes the mean-square error yields the final result.
\end{IEEEproof}

\section{Proof of Lemma \ref{lemma: expectation_bound}}

\begin{IEEEproof}
From the definition of the expectation of a bounded random variable $E_n=(X^*-\hat{X}_n)^2$:
\begin{align*}
		& \EE[(X^*-\hat{X}_n)^2] = \int_0^1 \PP((X^*-\hat{X}_n)^2>t) dt \\
			&= \int_0^{\Delta^2} \PP((X^*-\hat{X}_n)^2>t) dt + \int_{\Delta^2}^1 P((X^*-\hat{X}_n)^2>t) dt \\
			&\leq \Delta^2 + (1-\Delta^2) \PP(|X^*-\hat{X}_n|>\Delta).
\end{align*}
\end{IEEEproof}

\section{Proof of Theorem \ref{thm:ub}}

\begin{IEEEproof}
	Assume the pseudo-posterior after the $m$th player's response is $\ba^{(M-m)}(j+1)$, with the notation $\ba^{(0)}(j+1)=\ba(j+1)$. Let $k^*$ denote the index of the bin that contains $X^*$-i.e., $X^*\in I_{k^*}$. Define $M^{(m)}(j)= \frac{1}{a^{(M-m)}_{k^*}(j)} - 1$, with the notation $M^{(0)}(j)=M(j)$. Define the improvement ratio $N(j+1) = \frac{M(j+1)}{M(j)}$ for the $j$th time instant, and the improvement ratios $N^{(m)}(j+1)=\frac{M^{(m-1)}(j+1)}{M^{(m)}(j+1)}$ for $m=1,\dots,M-1$and $N^{(M)}(j+1)=\frac{M^{(M-1)}(j+1)}{M^{(0)}(j)}$. Let $\{\alpha_m\}_m$ denote the parameters associated with each player's pseudo-posterior update. By an application of Markov's inequality and repeated conditioning: 
	\begin{align*}
		\PP(&|X^*-X_n|>\Delta) \leq \PP(a_{k^*}(n)<1/2) \\
			&= \PP(M(n)>1) \leq \EE[M(n)] \\
			&= \EE[\EE[M(n-1)N(n)|\ba(n-1)]] \\
			&= \EE[M(n-1)\EE[N(n)|\ba(n-1)]] \\
			&= \cdots = M(0) \EE\left[ \prod_{l=1}^n \EE[N(l)|\ba(l-1)] \right] \\
			&\leq M(0) \left( \max_{0\leq j\leq n-1} \max_{\ba(j)} \EE[N(j+1)|\ba(j)] \right)^n.
	\end{align*}
	Theorem 1 in \cite{CastroNowak07} implies that after every discretized bisection $m$, we have the bound on the improvement ratio:
	\begin{equation} \label{eq:improvement_ratio}
		\EE[N^{(m)}(j+1)|\ba^{(m)}(j+1)] \leq \frac{1-\epsilon_m}{2(1-\alpha_m)}+\frac{\epsilon_m}{2\alpha_m} < 1
	\end{equation}
	Using the tower property of conditional expectations repeatedly again and using (\ref{eq:improvement_ratio}):
	\begin{align*}
		&\EE[N(j+1)|\ba(j)] = \EE\left[ \frac{M^{(0)}(j+1)}{M^{(0)}(j)} \Big| \ba^{(0)}(j) \right] \\
			&= \EE\left[ \frac{M^{(M-1)}(j+1)}{M^{(0)}(j)} \times \prod_{k=1}^{M-1} \frac{M^{(k-1)}(j+1)}{M^{(k)}(j+1)} \Big|\ba^{(0)}(j) \right] \\
			&= \EE\left[ \prod_{m=1}^{M} N^{(m)}(j+1) \Big|\ba^{(0)}(j) \right] \\
			&= \EE\left[ \EE[\prod_{m=1}^{M} N^{(m)}(j+1)\Big|\ba^{(1)}(j+1),\ba^{(0)}(j)] \Big|\ba^{(0)}(j) \right] \\
			&= \EE\left[ \prod_{m=2}^{M} N^{(m)}(j+1) \EE[N^{(1)}(j+1) \Big|\ba^{(1)}(j+1)] \Big|\ba^{(0)}(j) \right] \\
			&\leq \left(\frac{1-\epsilon_M}{2(1-\alpha_M)} + \frac{\epsilon_M}{2\alpha_M}\right) \EE\left[ \prod_{m=2}^{M} N^{(m)}(j+1) \Big|\ba^{(0)}(j) \right] \\
			&\leq \dots \leq \prod_{m=1}^M \left( \frac{1-\epsilon_m}{2(1-\alpha_m)}+\frac{\epsilon_m}{2\alpha_m} \right).
	\end{align*}
	To optimize the bound, we choose $\alpha_i = \frac{\sqrt{\epsilon_i}}{\sqrt{\epsilon_i}+\sqrt{1-\epsilon_i}},i=1,2$ to obtain:
	\begin{align*}
		\PP(|X^*-X_n|>\Delta) &\leq (\frac{1}{\Delta}-1) \left(\prod_{m=1}^M \Big(1-\bar{C}(\epsilon_m)\Big) \right)^n \\
			&\leq (\frac{1}{\Delta}-1) \exp \left( -n \sum_{m=1}^M \bar{C}(\epsilon_n) \right).
	\end{align*}
	This concludes the first part. The second part follows by applying Lemma \ref{lemma: expectation_bound}:
	\begin{equation*}
		\EE[(X^*-\hat{X}_n)^2] \leq \Delta^2 + \Delta^{-1} e^{-n\bar{C}}.
	\end{equation*}
	Optimizing the bound, we choose $\Delta = \Delta_n = 2^{-1/3} e^{-n\bar{C}/3}$, from which we conclude the second part.
\end{IEEEproof}

\section{Proof of Theorem \ref{thm:unknown_joint}}

\begin{IEEEproof}
	1) \underline{Optimality conditions} \\[0em]
	The solution of (\ref{eq: min_entr_obj}) yields the Bellman recursion \cite{Dynkin:1979}:
	\begin{equation*}
		V_n(p_n) = \inf_{\bA} \EE\left[V_{n+1}(p_{n+1})|\bA_n=\bA,\mathcal{F}_n\right]
	\end{equation*}
	Using a similar argument as in the proof of Theorem \ref{thm:joint}, the optimal solution at time $n$ is given by maximizing the entropy loss at time $n$:
	\begin{align*}
		G_n^* &= \sup_\bA I((X^*,\bepsilon^*); \bY_{n+1}|\bA_n=\bA,\mathcal{F}_n) \\
			&= \sup_{\bA} \left\{ H(p_n) - \EE\left[H(p_{n+1})|\bA_n=\bA,\mathcal{F}_n\right] \right\}
	\end{align*}
	and the value function is given by $V_n(p_n)=H(p_n)-\sum_{k=n}^{N-1} G_k^*$ for $n<N$ and $V_N(p_N)=H(p_N)$. We can expand the mutual information as:
	\begin{align*}
		I&((X^*,\bepsilon^*); \bY_{n+1}|\bA_n,\mathcal{F}_n) \\
		&= H(\bY_{n+1}|\bA_n,\mathcal{F}_n) - \EE\left[H(\bY_{n+1})|X^*,\bepsilon^*,\bA_n,\mathcal{F}_n\right]
	\end{align*}
	The conditional probability of $\bY_{n+1}$ given the query $\bA_n=\bA$ can be written as:
	\begin{align*}
		\PP&(\bY_{n+1}|\bA_n=\bA,\mathcal{F}_n) = \EE[\PP(\bY_{n+1}|\bA_n=\bA,X^*,\bepsilon^*,\mathcal{F}_n)] \\
			&= \int_{\bepsilon=0}^{1/2} \int_{x\in \mathcal{X}} \PP(\bY_{n+1}|\bA_n=\bA,X^*=x,\bepsilon^*=\bepsilon) p_n(x,\bepsilon) dx d\bepsilon \\
			&= \int_{\bepsilon=0}^{1/2} \int_{x\in \mathcal{X}} \Bigg(\prod_{m=1}^M f_1(Y_{n+1}^{(m)}|\epsilon_m) I(x\in A^{(m)}) \\
			&\qquad + f_0(Y_{n+1}^{(m)}|\epsilon_m) I(x\notin A^{(m)})  \Bigg) p_n(x,\bepsilon) dx d\bepsilon \\
			&= \int_{\bepsilon=0}^{1/2} \sum_{i_1:i_M=0}^1 g_{i_1:i_M}(\by|\bepsilon) \\
			&\qquad \times \left\{ \int_{x\in \mathcal{X}} I\left( \bigcap_{m} (A^{(m)})^{i_m} \right) p_n(x,\bepsilon) dx \right\} d\bepsilon \\
			&= \sum_{i_1:i_M=0}^1 \int_{\bepsilon=0}^{1/2} g_{i_1:i_M}(\by|\bepsilon) P_n\left(\bigcap_{m} (A^{(m)})^{i_m},\bepsilon \right) d\bepsilon
	\end{align*}
	where $p_n(x,\bepsilon)=p_n(x,\epsilon_1,\dots,\epsilon_M)$. This gives the first term in (\ref{eq:gain_joint}). To obtain the second term, notice:
	\begin{align*}
		&\EE[ H(\bY_{n+1})|X^*,\bepsilon^*,\bA_n=\bA,\mathcal{F}_n] \\
			&= \int_{\bepsilon} \int_{x\in \mathcal{X}} p_n(x,\bepsilon) H(\bY_{n+1}|X^*=x,\bepsilon^*=\bepsilon,\bA_n=\bA,\mathcal{F}_n) dx d\bepsilon \\
			&= \int_{\bepsilon} \left\{\sum_{i_1:i_M=0}^1 \int_{x\in \bigcap_m (A^{(m)})^{i_m}} p_n(x,\bepsilon) H(g_{i_1:i_M}(\cdot|\bepsilon)) dx \right\} d\bepsilon \\
			&= \sum_{i_1:i_M=0}^1 \int_{\bepsilon=0}^{1/2} H(g_{i_1:i_M}(\cdot|\bepsilon)) P_n\left(\bigcap_m (A^{(m)})^{i_m},\bepsilon\right)  d\bepsilon
	\end{align*}
	The proof is complete.\\[0em]
	2) \underline{Bounds on maximum entropy loss} \\[0em]
	First, we prove the upper bound. Note that the second term in (\ref{eq:gain_joint}) is independent of the queries, so the supremum can be restricted to only the first term without loss of generality. This is justified by using the additivity of the entropy of a product density:
	\begin{align*}
		H(&g_{i_1:i_M}(\cdot|\bepsilon)) = H\left( \prod_{m=1}^M f_{i_m}^{(m)}(\cdot|\epsilon_m) \right) \\
			&= \sum_{m=1}^M H(f_{i_m}^{(m)}(\cdot|\epsilon_m)) = \sum_{m=1}^M h_b(\epsilon_m)
	\end{align*}
	From part 1), the maximum entropy loss can be bounded from above as:
	\begin{align}
		&G_n^* \nonumber \\
		&= \sup_{\bA} H\left( \sum_{i_1:i_M=0}^1 \int_{\bepsilon=0}^{1/2} g_{i_1:i_M}(\cdot|\bepsilon) P_n\left(\bigcap_{m} (A^{(m)})^{i_m},\bepsilon\right) d\bepsilon \right) \nonumber \\
			&\quad - \sum_{i_1:i_M=0}^1 \int_{\bepsilon=0}^{1/2} H(g_{i_1:i_M}(\cdot|\bepsilon)) P_n\left(\bigcap_{m} (A^{(m)})^{i_m},\bepsilon\right) d\bepsilon \Big\} \nonumber \\
			&\leq \log_2(\card(\mathcal{Y})) - \int_{\bepsilon=0}^{1/2} \left\{ \sum_{m} h_B(\epsilon_m) \right\} \nonumber \\
			&\quad \times \left\{\sum_{i_1:i_M=0}^1 P_n\left(\bigcap_{m} (A^{(m)})^{i_m},\bepsilon\right) \right\} d\bepsilon  \label{eq:lemma_app} \\
			&= M - \sum_m \left\{\int_{\bepsilon_m=0}^{1/2} h_b(\epsilon_m) p_n(\epsilon_m) d\bepsilon\right\} \nonumber \\
			&= \sum_m (1-\EE[h_b(\epsilon_m)|\mathcal{F}_n]) \nonumber \\
			&= \EE\left[ \sum_m C(\epsilon_m) \Big| \mathcal{F}_n \right] \nonumber
	\end{align}
	where we used the fact that the capacity of a BSC is $C(\epsilon_m)=1-h_b(\epsilon_m)$. In (\ref{eq:lemma_app}), we also used the fact that the uniform distribution maximizes the entropy (see Ch.2 in \cite{CoverThomas}). 
	
	Second, we prove the lower bound. By the concavity of $H(\cdot)$, we obtain:
	\begin{align}
		&G_n^* \nonumber \\
		&= \sup_{\bA} \Big\{ H\left( \sum_{i_1:i_M=0}^1 \int_{\bepsilon=0}^{1/2} g_{i_1:i_M}(\cdot|\bepsilon) P_n\left(\bigcap_{m} (A^{(m)})^{i_m},\bepsilon\right) d\bepsilon \right) \nonumber \\
			&\quad - \sum_{i_1:i_M=0}^1 \int_{\bepsilon=0}^{1/2} H(g_{i_1:i_M}(\cdot|\bepsilon)) P_n\left(\bigcap_{m} (A^{(m)})^{i_m},\bepsilon\right) d\bepsilon \Big\} \nonumber \\
			&\geq \sup_{\bA} \Big\{ \int_{\bepsilon=0}^{1/2} H\left( \sum_{i_1:i_M=0}^1 g_{i_1:i_M}(\cdot|\bepsilon) P_n\left(\bigcap_{m} (A^{(m)})^{i_m} \Big| \bepsilon\right) \right) \nonumber\\
			&\qquad \times p_n(\bepsilon) d\bepsilon \nonumber \\
			&\quad -  \int_{\bepsilon=0}^{1/2} \sum_{i_1:i_M=0}^1 H(g_{i_1:i_M}(\cdot|\bepsilon)) P_n\left(\bigcap_{m} (A^{(m)})^{i_m}\Big|\bepsilon\right) p_n(\bepsilon) d\bepsilon \Big\} \nonumber \\
			&=  \sup_{A^{(1)},\dots,A^{(M)}} \EE \Big[H\left( \sum_{i_1:i_M=0}^1 g_{i_1:i_M}(\cdot|\bepsilon) P_n\left(\bigcap_{m} (A^{(m)})^{i_m} \Big| \bepsilon\right) \right)  \nonumber \\
			&\quad - \sum_{i_1:i_M=0}^1 H(g_{i_1:i_M}(\cdot|\bepsilon)) P_n\left(\bigcap_{m} (A^{(m)})^{i_m}\Big|\bepsilon\right) \Big|\mathcal{F}_n \Big] \nonumber 
	\end{align}
	\begin{align}
			&= \sup_{\bp:\bp \geq 0, 1^T\bp=1} \EE\Big[ H\left( \sum_{i_1:i_M=0}^1 g_{i_1:i_M}(\cdot|\bepsilon) p_{i_1,\dots,i_M} \right) \label{eq:sup_exp_interchange} \\
			&\quad - \sum_{i_1:i_M=0}^1 H(g_{i_1:i_M}(\cdot|\bepsilon)) P_n\left(\bigcap_{m} (A^{(m)})^{i_m}\Big|\bepsilon\right) \Big|\mathcal{F}_n \Big] \nonumber \\
			&= \EE\Big[ \sup_{\bp:\bp \geq 0, 1^T\bp=1} H\left( \sum_{i_1:i_M=0}^1 g_{i_1:i_M}(\cdot|\bepsilon) p_{i_1,\dots,i_M} \right) \nonumber \\
			&\quad - \sum_{i_1:i_M=0}^1 H(g_{i_1:i_M}(\cdot|\bepsilon)) p_{i_1,\dots,i_M} \Big|\mathcal{F}_n \Big] \nonumber \\
			&= \EE\left[\sum_{m=1}^M C(\epsilon_m) \Big| \mathcal{F}_n\right] \label{eq:cap}
	\end{align}
	where we used the consistent reparameterization $P_n\left(\bigcap_{m} (A^{(m)})^{i_m} \Big| \bepsilon\right) = p_{i_1,\dots,i_M}$ in (\ref{eq:sup_exp_interchange}) and Theorem \ref{thm:separation} in (\ref{eq:cap}).
\end{IEEEproof}

\section{Proof of Theorem \ref{thm:equivalence_unknown}}

\begin{IEEEproof}
	After querying all $M$ players in sequence, the entropy loss is: 
	\begin{align}
		G_{seq,n}^* &= \sup_{\{A_{n_t}\}_{t=0}^{M-1}} \EE[H(p_{n})-H(p_{n+1})|\mathcal{G}_n] \nonumber \\
			&= \sup_{\{A_{n_t}\}_{t=0}^{M-1}} \EE\left[\sum_{t=0}^{M-1} H(p_{n_t})-H(p_{n_{t+1}}) \Bigg| \mathcal{G}_n  \right] \label{eq:telescope} \\
			&= \sup_{\{A_{n_t}\}_{t=0}^{M-1}} \sum_{t=0}^{M-1} \EE\left[ \EE\left[ H(p_{n_t})-H(p_{n_{t+1}}) \Bigg|A_{n_t}, \mathcal{G}_{n_t} \right] \Bigg| \mathcal{G}_n \right] \label{eq:tower} \\
			&= \sup_{\{A_{n_t}\}_{t=0}^{M-1}} \EE\left[ \sum_{t=0}^{M-1} I((X^*,\bepsilon^*);Y_{n_{t+1}}|A_{n_t},\mathcal{G}_{n_t}) \Bigg| \mathcal{G}_n \right] \label{eq:MI} \\
			&= \EE\left[ \sum_{t=0}^{M-1} \sup_{A_{n_t}} I((X^*,\bepsilon^*);Y_{n_{t+1}}|A_{n_t},\mathcal{G}_{n_t}) \Bigg| \mathcal{G}_n  \right] \nonumber \\
			&= \EE\left[ \sum_{t=0}^{M-1} C(\epsilon_{t+1}) \Bigg| \mathcal{G}_n \right] = \EE\left[ \sum_{m=1}^M C(\epsilon_m) \Bigg| \mathcal{G}_n \right] \nonumber
	\end{align}
	where we used a telescoping sum in (\ref{eq:telescope}) and the tower property of expectation with $\mathcal{G}_{n_t} \supseteq \mathcal{G}_n$ in (\ref{eq:tower}). In (\ref{eq:MI}), we used the optimality condition of maximum entropy loss by applying Theorem \ref{thm:unknown_joint} with $M=1$ for each sub-instant $n_t$ with $m=t+1$:
	\begin{align*}
		\sup_{A_{n_t}} &\left\{ H(p_{n_t}) - \EE[H(p_{n_{t+1}})|A_{n_t},\mathcal{G}_{n_t}] \right\} \\
		&= \sup_{A_{n_t}} I((X^*,\bepsilon^*);Y_{n_{t+1}}|A_{n_t},\mathcal{G}_{n_t}) \\
		&= \EE[ C(\epsilon_{t+1}) | \mathcal{G}_{n_t} ]
	\end{align*}
	The second part follows from Theorem \ref{thm:unknown_joint} part 2).
\end{IEEEproof}

\section{Proof of Theorem \ref{thm:unknown}}

\begin{IEEEproof}
	The solution of (\ref{eq: min_entr_obj}) yields the Bellman recursion:
	\begin{equation*}
		V_n(p_n) = \inf_{u,A} \EE\left[V_{n+1}(p_{n+1})|u_n=u,A_n=A,\mathcal{F}_n\right]
	\end{equation*}
	Using a similar argument as in Theorem 2 in \cite{Jedynak12}, the optimal solution at time $n$ is given by maximizing the entropy loss at time $n$:
	\begin{align*}
		G_n &= \max_u \sup_A I((X^*,\bepsilon^*); Y_{n+1}^{(u)}|u_n=u,A_n^{(u)}=A,\mathcal{F}_n) \\
		&= H(p_n) - \EE\left[H(p_{n+1})|u_n=u,A_n^{(u)}=A,\mathcal{F}_n\right]
	\end{align*}
	and the value function is given by $V_n(p_n)=H(p_n)-\sum_{k=n}^{N-1} G_k$ for $n<N$ and $V_N(p_N)=H(p_N)$. We can expand the mutual information:
	\begin{align*}
		I&((X^*,\bepsilon^*); Y_{n+1}^{(u)}|u_n,A_n^{(u)},\mathcal{F}_n) \\
		&= H(Y_{n+1}^{(u)}|u_n,A_n^{(u)},\mathcal{F}_n) - \EE\left[H(Y_{n+1}^{(u)})|X^*,\bepsilon^*,u_n,A_n^{(u)},\mathcal{F}_n\right]
	\end{align*}
	The conditional probability of $Y_{n+1}^{(u)}$ given the selection $u_n=u$ and the query $A_n^{(u)}=A$:
	\begin{align*}
		&\PP(Y_{n+1}^{(u)}|u_n=u,A_n^{(u)}=A,\mathcal{F}_n) \\
			&= \EE[\PP(Y_{n+1}^{(u)}|u_n=u,A_n^{(u)}=A,X^*,\bepsilon^*,\mathcal{F}_n)] \\
			&= \int_{\bepsilon} \int_{x\in \mathcal{X}} \left( f_1(Y_{n+1}^{(u)}|\epsilon_u) I(x\in A) + f_0(Y_{n+1}^{(u)}|\epsilon_u) I(x\notin A)  \right) \\
			&\qquad \times p_n(x,\bepsilon) dx d\bepsilon \\
			&= \int_{\epsilon_u=0}^{1/2} \int_{x\in \mathcal{X}} \left( f_1(Y_{n+1}^{(u)}|\epsilon_u) I(x\in A) + f_0(Y_{n+1}^{(u)}|\epsilon_u) I(x\notin A)  \right) \\
			&\qquad \times p_n^{(u)}(x,\epsilon_u) dx d\epsilon_u \\
			&= \int_{\epsilon_u=0}^{1/2} f_1(Y_{n+1}^{(u)}|\epsilon_u) P_n^{(u)}(A,\epsilon_u) + f_0(Y_{n+1}^{(u)}|\epsilon_u) P_n^{(u)}(A^c,\epsilon_u) d\epsilon_u
	\end{align*}
	where $p_n^{(u)}(x,\epsilon_u)=\int_{\{\epsilon_m\in[0,1/2):m\neq u\}} p_n(x,\bepsilon) d\{\epsilon_m:m\neq u\}$ denotes the $u$th sub-marginal. This gives the first term in (\ref{eq:gain}). To obtain the second term, notice:
	\begin{align*}
		&\EE[H(Y_{n+1}^{(u)})|X^*,\bepsilon^*,u_n=u,A_n^{(u)}=A,\mathcal{F}_n] \\
			&= \int_{\bepsilon} \int_{x\in \mathcal{X}} H(Y_{n+1}^{(u)}|X^*=x,\bepsilon^*=\bepsilon,u_n=u,A_n^u=A,\mathcal{F}_n) \\
			&\qquad \times p_n(x,\bepsilon) dx d\bepsilon \\
			&= \int_{\bepsilon} \Bigg\{\int_{x\in A} p_n(x,\bepsilon) H(f_1(Y_{n+1}^{(u)}|\epsilon_u)) dx \\
			&\qquad + \int_{x\notin A} p_n(x,\bepsilon) H(f_0(Y_{n+1}^{(u)}|\epsilon_u)) dx \Bigg\} d\bepsilon \\
			&= \int_{\epsilon_u=0}^{1/2} H(f_1(\cdot|\epsilon_u)) P_n^{(u)}(A,\epsilon_u) \\
			&\qquad + H(f_0(\cdot|\epsilon_u)) P_n^{(u)}(A^c,\epsilon_u) d\epsilon_u
	\end{align*}
	The proof of the first part is complete. The second part follows from part (2) of Theorem \ref{thm:unknown_joint}.
\end{IEEEproof}

\section{Proof of Corollary \ref{cor:unknown}}

\begin{IEEEproof}
	From Theorem \ref{thm:unknown}, we have the optimality condition shown in (\ref{eq:gain}). Under Assumption \ref{assump:BSC}, we have $H(f_0(\cdot|\epsilon_u)=H(f_1(\cdot|\epsilon_u))=h_B(\epsilon_u)$. Using this in the second term in the supremum of (\ref{eq:gain}):
	\begin{align}
		&\int_{\epsilon_u=0}^{1/2} H\left(f_1(\cdot|\epsilon_u)\right) P_n^{(u)}(A,\epsilon_u) + H\left(f_0(\cdot|\epsilon_u)\right) P_n^{(u)}(A^c,\epsilon_u) d\epsilon_u \nonumber \\
			&\quad = \int_{\epsilon_u=0}^{1/2} h_B(\epsilon_u) \left(P_n^{(u)}(A,\epsilon_u) + P_n^{(u)}(A^c,\epsilon_u)\right) d\epsilon_u \nonumber \\
			&\quad = \int_{\epsilon_u=0}^{1/2} h_B(\epsilon_u) p_n^{(u)}(\epsilon_u) d\epsilon_u = c_n^{(u)} \label{eq:term2}
	\end{align}
	Thus, we conclude that the second term in (\ref{eq:gain}) is independent of the query region $A$, but still depends on the player $u$.
	
	Rewriting the first term in the supremum of (\ref{eq:gain}), we have for $A=[0,x]$:
	\begin{align}
		H&\left( \int_{\epsilon_u=0}^{1/2} f_1(\cdot|\epsilon_u) P_n^{(u)}(A,\epsilon_u) + f_0(\cdot|\epsilon_u) P_n^{(u)}(A^c,\epsilon_u) d\epsilon_u \right) \nonumber \\
			&= H\Bigg( \int_{\epsilon_u=0}^{1/2} f_1(\cdot|\epsilon_u) \left\{\int_0^x p_n^{(u)}(t,\epsilon_u) dt \right\} \nonumber \\
			&\qquad + f_0(\cdot|\epsilon_u) \left\{ \int_x^1 p_n^{(u)}(t,\epsilon_u) dt \right\}  d\epsilon_u \Bigg) \nonumber \\
			&= H\Bigg( \int_0^x \left\{ \int_{\epsilon_u=0}^{1/2} f_1(\cdot|\epsilon_u) p_n^{(u)}(t,\epsilon_u) d\epsilon_u \right\} dt  \nonumber \\
			&\qquad + \int_x^1  \left\{ \int_{\epsilon_u=0}^{1/2} f_0(\cdot|\epsilon_u) p_n^{(u)}(t,\epsilon_u) d\epsilon_u \right\} dt \Bigg) \nonumber \\
			&= h_B(g_{1,n}^{(u)}(x)) \label{eq:term1}
 	\end{align}
	where $g_{1,n}^{(u)}(x)$ is defined in the statement of the theorem.
\end{IEEEproof}


%


\bibliographystyle{IEEEtran}
\bibliography{refs}

\newpage
\begin{IEEEbiographynophoto}
{Theodoros Tsiligkaridis} received his B.Sc. degree (cum laude) in electrical engineering from the University of Washington (UW), Seattle, in 2008. He received the M.Sc. degree in electrical and computer engineering from the University of Illinois at Urbana-Champaign (UIUC) in 2009, under an ECE Dept. Distinguished Fellowship. He completed his Ph.D. in electrical engineering and computer science at the University of Michigan (UMich), Ann Arbor, under a Rackham Engineering Awards Fellowship.

Dr. Tsiligkaridis is currently a member of the Technical Research Staff at MIT Lincoln Laboratory, Lexington, in the Advanced Sensor Techniques Group. His research interests include statistical signal processing, network science, controlled sensing, information theory and applications.

\end{IEEEbiographynophoto}
\vfill
\begin{IEEEbiographynophoto}
{Brian M. Sadler} received the B.S. and M.S. degrees from the University of Maryland, College Park, and the PhD degree from the University of Virginia, Charlottesville, all in electrical engineering.  He is a Fellow of the Army Research Laboratory (ARL) in Adelphi, MD.  

Dr. Sadler is an associate editor for EURASIP Signal Processing, was an associate editor for the IEEE Transactions on Signal Processing and IEEE Signal Processing Letters, and has been a guest editor for several journals including IEEE JSTSP, IEEE JSAC, the IEEE SP Magazine, and the International Journal of Robotics Research. He is a member of the IEEE Signal Processing Society Sensor Array and Multi-channel Technical Committee, and Co-Chair of the IEEE Robotics and Automation Society Technical Committee on Networked Robotics.  He received Best Paper Awards from the Signal Processing Society in 2006 and 2010.  He has received several ARL and Army R\&D awards, as well as a 2008 Outstanding Invention of the Year Award from the University of Maryland.  His research interests include information science, networked and autonomous systems, sensing, and mixed-signal integrated circuit architectures.

\end{IEEEbiographynophoto}
\vfill
\begin{IEEEbiographynophoto}
{Alfred O. Hero III}
received the B.S. (summa cum laude) from Boston University (1980) and the Ph.D from Princeton University (1984), both in Electrical Engineering. Since 1984 he has been with the University of Michigan, Ann Arbor, where he is the R. Jamison and Betty Williams Professor of Engineering. His primary appointment is in the Department of Electrical Engineering and Computer Science and he also has appointments, by courtesy, in the Department of Biomedical Engineering and the Department of Statistics. From 2008 to 2013 he held the Digiteo Chaire d'Excellence, sponsored by Digiteo Research Park in Paris, located at the Ecole Superieure d'Electricite, Gif-sur-Yvette, France. He has held other visiting positions at LIDS Massachusetts Institute of Technology (2006), Boston University (2006), I3S University of Nice, Sophia-Antipolis, France (2001), Ecole Normale Sup\'erieure de Lyon (1999), Ecole Nationale Sup\'erieure des T\'el\'ecommunications, Paris (1999), Lucent Bell Laboratories (1999), Scientific Research Labs of the Ford Motor Company, Dearborn, Michigan (1993), Ecole Nationale Superieure des Techniques Avancees (ENSTA), Ecole Superieure d'Electricite, Paris (1990), and M.I.T. Lincoln Laboratory (1987 - 1989).

Alfred Hero is a Fellow of the Institute of Electrical and Electronics Engineers (IEEE). He received the University of Michigan Distinguished Faculty Achievement Award (2011). He has been plenary and keynote speaker at several workshops and conferences. He has received several best paper awards including: an IEEE Signal Processing Society Best Paper Award (1998), a Best Original Paper Award from the Journal of Flow Cytometry (2008), a Best Magazine Paper Award from the IEEE Signal Processing Society (2010), a SPIE Best Student Paper Award (2011), an IEEE ICASSP Best Student Paper Award (2011), an AISTATS Notable Paper Award (2013), and an IEEE ICIP Best Paper Award (2013). He received an IEEE Signal Processing Society Meritorious Service Award (1998), an IEEE Third Millenium Medal (2000), an IEEE Signal Processing Society Distinguished Lecturership (2002), and an IEEE Signal Processing Society Technical Achievement Award (2014). He was President of the IEEE Signal Processing Society (2006-2007). He was a member of the IEEE TAB Society Review Committee (2009), the IEEE Awards Committee (2010-2011), and served on the Board of Directors of the IEEE (2009-2011) as Director of Division IX (Signals and Applications). He presently serves on the IEEE TAB Nominations and Appointments Committee.

Alfred Hero's recent research interests are in statistical signal processing, machine learning and the analysis of high dimensional spatio-temporal data. Of particular interest are applications to networks, including social networks, multi-modal sensing and tracking, database indexing and retrieval, imaging, biomedical signal processing, and biomolecular signal processing.

\end{IEEEbiographynophoto}

\end{document}